\documentclass[12pt]{article}
\usepackage{a4}
\usepackage{amsfonts}
\usepackage{amssymb}
\usepackage{cite}
\usepackage{epsfig}
\usepackage{amsmath}

\usepackage[english]{babel}

\author{}

\newcommand{\drawsquare}[2]{\hbox{%
\rule{#2pt}{#1pt}\hskip-#2pt
\rule{#1pt}{#2pt}\hskip-#1pt
\rule[#1pt]{#1pt}{#2pt}}\rule[#1pt]{#2pt}{#2pt}\hskip-#2pt
\rule{#2pt}{#1pt}}
\newcommand{\Ysymm}{\raisebox{-.5pt}{\drawsquare{6.5}{0.4}}\hskip-0.4pt%
         \raisebox{-.5pt}{\drawsquare{6.5}{0.4}}}
\newcommand{\Yasymm}{\raisebox{-3.5pt}{\drawsquare{6.5}{0.4}}\hskip-6.9pt%
        \raisebox{3pt}{\drawsquare{6.5}{0.4}}}

\newcommand{\be}{\begin{equation}}
\newcommand{\ee}{\end{equation}}
\newcommand{\ba}{\begin{array}}
\newcommand{\ea}{\end{array}}
\newcommand{\bea}{\begin{eqnarray}}
\newcommand{\eea}{\end{eqnarray}}

\def\IR{\relax{\rm I\kern-.18em R}}

\def\IP{\relax{\rm I\kern-.18em P}}
\def\inbar{\vrule height1.5ex width.4pt depth0pt}
\def\IC{\relax\,\hbox{$\inbar\kern-.3em{\rm C}$}}

\newcommand{\Z}{\mathbb{Z}}

\def\K3{{\bf K3}}

\def\n2d{\cN_{V^*}^{\otimes 2}}

\def\IC{\mathbb{C}}

\def\IR{\mathbb{R}}

\def\IP{\mathbb{P}}

\def\cN{{\mathcal N}}

\def\beq{\begin{equation}}
\def\eeq{\end{equation}}
\def\beqa{\begin{eqnarray}}
\def\eeqa{\end{eqnarray}}

\begin{document}

\title{
\begin{flushright} \vspace{-2cm}
{\small IFT-UAM/CSIC-08-74\\
\small UPR-1202-T\\
} \end{flushright} \vspace{4.0cm}
 Stringy Instantons and Yukawa Couplings\\
in MSSM-like Orientifold Models}
 \vspace{1.0cm}
\author{\small  L.E. Ib\'a\~nez$^1$  and R. Richter$^2$}

\date{}

\maketitle

\begin{center}
\emph{${}^1$ Departamento de Fisica Te\'orica
and Instituto de Fisica Te\'orica UAM-CSIC,\\
Universidad Aut\'onoma de Madrid, Cantoblanco, 28049 Madrid, Spain}
\emph{$^{2}$Department of Physics and Astronomy, University of Pennsylvania, \\
     Philadelphia, PA 19104-6396, USA }
\vspace{0.2cm}

\tt{ ibanez@madriz1.ft.uam.es, rrichter@physics.upenn.edu }
\vspace{1.0cm}
\end{center}
\vspace{0.7cm}

\begin{abstract}

Type IIA  orientifold constructions with intersecting D6-branes and
their IIB duals in terms of magnetized D9/D7-branes constitute one
of the most promising avenues for the construction of semirealistic
MSSM-like compactifications. One generic problem with these
constructions is, that there are many Yukawa couplings, which vanish
due to additional $U(1)$ symmetries in the theory. In this paper we
consider a number of such settings and study, under what conditions
stringy instanton effects can give rise to non-perturbative
contributions to the Yukawa couplings, so that all perturbatively
forbidden terms are induced. We find specific settings, in which
indeed Yukawa couplings for all fermions are obtained. For some cases we provide specific local examples of rigid
$O(1)$ instantons within the $T^6/{\mathbb Z}_2 \times {\mathbb
Z}_2'$ toroidal orientifold with torsion, giving rise to the required
amplitudes. A potential problem in these settings is, that the same
instantons, providing for Yukawa coupling contributions, may give rise
to too large $\mu$-terms for the Higgs multiplets. We show how this
problem may be overcome in explicit models with a doubled Higgs
system.

\end{abstract}

\thispagestyle{empty} \clearpage


\section{Introduction}

Due to their very appealing geometrical interpretation intersecting
brane worlds (for recent reviews on this subject see
\cite{Blumenhagen:2005mu,Blumenhagen:2006ci,Marchesano:2007de}) have
been a popular playground for realistic model building. In such
models the gauge groups appear on stacks of D6-branes, filling out
the four-dimensional spacetime and wrapping three-cycles in the
six-dimensional internal compactification manifold. Chiral matter
arises at intersections of two stacks of D6-branes, wrapping
different three-cycles in the internal manifold and their
multiplicity is encoded in the intersection number of the respective
three-cycles.

Over the last decade many intersecting brane models, giving rise to
MSSM- and GUT-like spectrum, have been constructed using mostly
toroidal orbifolds as compactification manifold\footnote{For
original work on non-supersymmetric intersecting D-branes, see
\cite{Blumenhagen:2000wh,Aldazabal:2000dg,Aldazabal:2000cn,Blumenhagen:2001te}
and for chiral supersymmetric ones see
\cite{Cvetic:2001tj,Cvetic:2001nr}.}. Given these, the next question
is, if we can reproduce finer details of the MSSM, such as Yukawa
couplings as well as their hierarchies. For intersecting D-brane
constructions the Yukawa couplings for chiral matter fields can be
extracted from string amplitudes
\cite{Abel:2003vv,Cvetic:2003ch,Lust:2004cx}. These are suppressed
by the open string world-sheet instantons connecting the three
intersecting branes \cite{Aldazabal:2000cn,Cremades:2003qj}. The
latter potentially give rise to interesting Yukawa hierarchies,
which, in principle, could reproduce the quark mass matrices
observed in the MSSM.

However, for specific intersecting D-brane constructions on toroidal
backgrounds, one often observes a factorization in terms of family
indices of the Yukawa couplings. This is happens for instance if
non-trivial intersections appear in different two tori of the
orientifold \cite{Cremades:2003qj}. The factorization of the Yukawa
couplings for the different matter fields (u-quark, d-quark,
electron and neutrino) allows only one of the three families to
acquire dirac masses. While this might explain the hierarchy between
the masses of the heaviest family compared to the two lightest
family \cite{Cremades:2003qj}, one still faces the problem, how to
generate masses for the other families. As demonstrated in
\cite{Cvetic:2002wh} (see also
\cite{Chen:2007px,Chen:2007zu,Chen:2008rx}) additional Higgs pairs
might overcome this issue and for appropriate values of the open and
closed string moduli give rise to desired Yukawa hierarchies. Let us
emphasize that for constructions on more general backgrounds such
factorizations of Yukawa couplings is expected to be absent.

Nevertheless, in the existing constructions typically some of the
Yukawa couplings are forbidden due to the violation of  global
$U(1)$'s, ruling them out of being realistic. These global $U(1)$'s
are remnants of the Green-Schwarz mechanism, a necessary ingredient
to ensure the cancellation of pure abelian as well as mixed
anomalies.

Recently, it has been realized that D-brane instantons can break
these global $U(1)$'s and induce otherwise forbidden couplings
\cite{Blumenhagen:2006xt,Haack:2006cy,Ibanez:2006da,Florea:2006si}.
These non-perturbative effects cannot be described as gauge instantons
in field theory, thus are purely stringy. For type IIA
compactification the relevant class of instantons are, so called,
$E2$-instantons, wrapping a three-cycle in the internal manifold and
being localized in four-dimensional spacetime. Under a $U(1)_a$
gauge transformation the action of such an $E2$-instanton, wrapping
the three-cycle $\pi_{E2}$, transforms as

\begin{align}
e^{-S_{E2}}=\exp\left[ \frac{2\pi}{ \ell_s^3}
           \left( -\frac{1}{g_s} {\rm Vol}_{\pi_{E2}} + i \int_{\pi_{E2}} C^{(3)}
         \right) \right] \longrightarrow  e^{i\, Q_a(E2)\,\Lambda_a}  \,\, e^{-S_{E2}}
\end{align} with\footnote{Note that, in contrast to \cite{Blumenhagen:2006xt}, there
is an additional minus sign in \eqref{charge of the instanton},
which is due to the fact that a positive intersection number
$I_{E2a}$ corresponds to the transformation behavior
$(E2,\overline{a})$ of the charged fermionic zero modes, rather than
$(\overline{E2},a)$. In the sequel we denote the real part of
$S_{E2}$, the instanton suppression factor, $S^{cl}_{E2}$.}
\begin{align}
 Q_a(E2) = -N_a \,\, \pi_{E2} \circ \left[\pi_a
-\pi'_a\right] \,\,.
 \label{charge of the instanton}\end{align}
Here $\pi_a$ and $\pi_{a'}$ denote the three-cycles which the brane
$a$ and its orientifold image $a'$ wrap and $N_a$ is the number
of $D6$-branes for stack $a$. A  superpotential term can be induced
if the product
\begin{align}
W^{np}= \prod_{i} \Phi_{a_i\,b_i} e^{-S_{E2}}
\end{align}
is invariant under all U(1) gauge transformations, thus if the $E2$
instanton compensates for all global U(1)'s carried just by the
product $\prod_{i} \Phi_{a_i\,b_i}$. Due to their non-perturbative
nature these couplings are suppressed, which potentially gives an
explanation for various hierarchies observed in nature.

Various applications of these novel non-perturbative effects in
different branches of the string landscape have  appeared
\cite{Blumenhagen:2006xt,Haack:2006cy,Ibanez:2006da,Florea:2006si,Abel:2006yk,
Akerblom:2006hx,Bianchi:2007fx,Cvetic:2007ku,Argurio:2007qk,Argurio:2007vq,
Bianchi:2007wy,Ibanez:2007rs,Akerblom:2007uc,Antusch:2007jd,Blumenhagen:2007zk,
Aharony:2007pr,Aharony:2007db,Blumenhagen:2007bn,Billo:2007sw,Billo:2007py,
Aganagic:2007py,Camara:2007dy,Cvetic:2007qj,Ibanez:2007tu,GarciaEtxebarria:2007zv,
Petersson:2007sc,Blumenhagen:2007sm,Bianchi:2007rb,Blumenhagen:2008ji,Matsuo:2008nu,Argurio:2008jm,Cvetic:2008ws,
Cvetic:2008hi,GarciaEtxebarria:2008pi,Buican:2008qe,Forcella:2008au,Blumenhagen:2008kq,Camara:2008zk,Billo':2008sp,Cvetic:2008mh,
Billo':2008pg,Kumar:2008cm,Marsano:2008jq,Marsano:2008py,Uranga:2008nh,Heckman:2008qt,GarciaEtxebarria:2008iw,Jelinski:2008ka}
\footnote{For two reviews on the subject of novel stringy instanton
effects see \cite{Akerblom:2007nh,Cvetic:2007sj}. For related
earlier work see \cite{Witten:1996bn,Ganor:1996pe,Billo:2002hm}.}.
These include the explicit generation of  perturbatively forbidden
couplings, such as Majorana masses for the right-handed Neutrinos,
$\mu$-terms for the MSSM Higgs sector or Yukawa-couplings of type
${\bf 10}\cdot{\bf 10}\cdot{\bf 5}_H$ in $SU(5)$-like GUT models.
Furthermore, stringy instanton effects play a crucial role in the
study of dynamical supersymmetry breaking, as well as moduli
stabilization. Other, more formal aspects, involve lifting of
additional zero modes via fluxes, multiinstanton contributions as
well as global issues of these non-perturbative effects.

In this work, we analyze various D-brane constructions presented in
\cite{Ibanez:2001nd} and \cite{Cremades:2002cs}, which give rise to
the MSSM spectrum or extensions of it, with respect to their Yukawa
couplings
\footnote{ In the alternative MSSM-like model of ref.\cite{Cremades:2002qm,Cremades:2003qj,
Marchesano:2004xz}
there are no Yukawa couplings  forbidden by any $U(1)$ symmetries
and hence charged instantons cannot induce any extra Yukawa coupling.}
 . In case the desired couplings are perturbatively forbidden,
since they violate global $U(1)$ selection rules, we present
conditions under which these couplings can be generated via
E2-instantons. We investigate the implications of these
non-perturbative effects on the four-dimensional phenomenology and
compare them to the observed experimental structure. Furthermore, we
present local realizations on the $T^6/{\mathbb Z}_2 \times {\mathbb
Z}_2'$ orbifold with discrete torsion \cite{Blumenhagen:2005tn}, in
which various couplings are generated by rigid $O(1)$ instantons. We
demonstrate for these local models that the instanton indeed have
the right suppression factor to yield to results compatible with
experiments.

Let us point out that the derived results also apply to the T-dual
Type I framework, where the superpotential receives contributions
from $E1$ and $E5$ instantons. This corner of the string landscape
is amenable to algebraic geometry techniques and has proven to be
more promising for constructions of globally consistent
semi-realistic string vacua, which exhibit stringy instanton
corrections to the superpotential
\cite{Cvetic:2007qj,Cvetic:2008mh}. The $O(1)$ instantons in Type
IIA correspond to $E1$ instantons in Type I. Thus all the conditions
on the $E2$-instantons presented in the course of this work can be
easily translated into the Type I framework, where the Yukawa
coupling inducing instanton is a rigid $E1$ instanton.


\section{E2-instanton\label{review}}

As pointed out in the introduction, $E2$-instantons with the right
charge under the global $U(1)$ can generate a superpotential term of
the form
\begin{align}
W^{np}= \prod_{i} \Phi_{a_i\,b_i} e^{-S_{E2}}\,\,,
\label{npSuperpotential}
\end{align}
which is otherwise forbidden. Note that the condition of carrying
the right charge is just a necessary one, but not sufficient.
Whether  an instanton indeed contributes to the superpotential,
depends crucially on its zero mode structure. After briefly
recalling the zero mode structure for a generic instanton
\cite{Cvetic:2007ku,Ibanez:2007rs}, we review the computation of a
non-perturbative superpotential contribution, induced by a rigid
$O(1)$ instanton \cite{Blumenhagen:2006xt}.

An $E2$ instanton, wrapping a generic three-cycle $\pi_{E2}$, gives
rise to the four bosonic zero modes $x^{\mu}$ due to the breakdown
of four-dimensional Poincar{\'e} invariance. They are accompanied
with four fermionic zero modes $\theta^{\alpha}$ and
$\overline{\tau}^{\dot{\alpha}}$ indicating the breakdown of the ${
\cal N}=2 $ supersymmetry, preserved by the Calabi-Yau manifold,
down to ${\cal N}=1$ supersymmetry. In addition the instanton $E2$
exhibits $b_1(\pi_{E2})$ zero modes associated with deformations of
the special Lagrangian cycle $\pi_{E2}$. Moreover, in the presence
of multiple instantons there appear zero modes at intersections of
two instantons.

Finally, there are fermionic zero modes charged under the $D6$
branes, which we call charged fermionic zero modes in the sequel.
These arise at intersections between the instanton $E2$ and the
D6-branes. Due to the GSO-projection only the chiral fermion is
present, which is crucial for the holomorphicity of the
superpotential. Adding up the D6-brane charge of all charged
fermionic zero modes, gives the total charge of the instanton
$Q_a(E2)$, which coincides with \eqref{charge of the instanton}.

The non-perturbative contribution is given by the path integral over
all instanton zero modes, thus in order to give rise to F-terms we
expect all uncharged zero modes, apart from $x^{\mu}$, and
$\theta^{\alpha}$, to be projected out or lifted. There are various
ways to ensure the absence of these additional undesired zero modes,
such as lifting via
fluxes\cite{Blumenhagen:2007bn,Billo':2008sp,Billo':2008pg,Uranga:2008nh},
via additional instantons
\cite{Blumenhagen:2007bn,GarciaEtxebarria:2007zv,Cvetic:2008ws,GarciaEtxebarria:2008pi}
or via additional interaction terms in case the instanton wraps a
cycle which coincides with one of the spacetime filling D6-branes
\cite{Akerblom:2006hx,Petersson:2007sc}.

Here we focus on a class of instantons wrapping a rigid, orientifold
invariant cycle in the internal manifold, so called rigid $O(1)$
instantons. For those the undesired $\overline{\tau}^{\dot{\alpha}}$
modes get projected out \cite{Florea:2006si,Argurio:2007vq,
Bianchi:2007wy,Ibanez:2007rs} and due to the rigidity of the
three-cycle there are no zero modes associated with the deformation
of the three-cycle. Moreover, in that case the $U(1)_a$ charge of
the instanton $E2$ simplifies to
\begin{align}
 Q_a(E2) = - N_a \,\, \pi_{E2} \circ \pi_a
 \label{charge of the O(1) instanton}\end{align}
due to the identification of the $E2-a$ and $E2-a'$ sector. Then the
pathintegral takes the form
\begin{align}
W^{np}= \int d^4x\, d^2 \theta \prod_i d\lambda_{a_i} \,
e^{-S^{cl}_{E2}} \, e^{-S^{int}}
\end{align}
Here $e^{-S^{cl}_{E2}}$ is the suppression factor\footnote{The
suppression factor $e^{-S^{cl}_{E2}}$ is the real part of $S_{E2}$
in \eqref{npSuperpotential}.} and $S^{int}$ stands for all
interaction terms involving the charged instanton zero modes
$\lambda_{a_i}$ and matter fields $\Phi$. Due to holomorphicity of
the superpotential, $S^{int}$ contains only disk amplitudes,
carrying exactly two charged zero modes $\lambda$, and annulus
contributions with no charged zero modes inserted
\cite{Blumenhagen:2006xt}. The latter contribution has been
calculated in \cite{Abel:2006yk,Akerblom:2006hx,Akerblom:2007uc} and
is related to the regularized threshold correction to the gauge
coupling of a $D6$-brane wrapping the same cycle as the instanton.
Thus the computation of the non-perturbative superpotential
contribution results into the calculation of various disk diagrams
involving the charged zero modes.

\section{Four-stack models\label{four stack models}}

A very natural way of realizing the MSSM is to embed the matter
content at intersections of 4 stacks of $D6$-branes giving rise to
the gauge symmetry
\begin{align}
U(3)_a\times U(2)_b \times U(1)_c \times U(1)_d\,\,.
\end{align}
The left-handed quarks $q_{L}$ are localized at the intersection of
brane $a$ and $b$ or its orientifold image $b'$, while the
right-handed quarks, $u_R$ and $d_R$, arise at intersections of
brane $a$ with one of the $U(1)$ branes and its orientifold image.
The left-handed leptons are charged under the $U(2)$ and neutral under $U(3)$,
thus appear at intersections between brane $b$ and one of the $U(1)$
branes. Finally, the right-handed electron $E_R$ and the
right-handed neutrino $N_R$, both singlet under $U(3)$ and $U(2)$
arise at intersections of two $U(1)$ branes.

Tadpole cancellation, essential for global consistency, requires
equal number of fields transforming as fundamental, $a$, and  as
anti-fundamental, $\overline{a}$, under the gauge group
$U(N_a)$\footnote{Here we assume the absence of symmetrics and
anti-symmetrics under $U(N_a)$}. This hold also true $U(2)$, whose
fundamental representation is real, thus equal to the
anti-fundamental one.

For a subclass of 4-stack models the constraint, mentioned above,
can be satisfied within the MSSM matter content.  This can be
achieved by requiring that exactly 2 generations of the left-handed
quarks $q_L$ arise from the sector $ab'$, while the third family is
localized at the intersection of brane $a$ and $b$. Ensuring that
the number of fundamentals is equal the number of anti-fundamentals
puts additional constraints on the transformation behavior of the
matter and Higgs fields. Even though, the homology classes of the
D-branes $a$, $b$, $c$ and $d$ alone might not cancel all the
tadpoles, one may  assume that one can find  additional D-branes
ensuring global consistency, in such a way, that they do not give
rise to any additional chiral exotics charged under the matter
$D6$-branes. For such local setups it is possible to investigate
various phenomenological aspects without knowing the details of the
global realization.

In this paper we discuss three different setups which give rise to
the MSSM and extensions of it. We analyze which Yukawa couplings are
perturbatively realized and investigate under what circumstances an
instanton induces the perturbatively missing, but desired couplings.
Let us list for all three setups the matter, charged under $SU(3)$
and $SU(2)$

\begin{itemize}
\item[(1)]$1\times Q_l=(3, \overline 2)$ \,\,\,\,$2 \times q_l=(3,2)$ \,\,\,\,
$3 \times l=(1,\overline{2})$\\ \\ $ 1 \times (H_d + H_u)^{N=2}=
(1,\overline 2)+(1, 2)$
\item[(2)] $1\times Q_l=(3, \overline 2)$ \,\,\,\,$2 \times q_l=(3,2)$ \,\,\,\,
$2 \times l=(1,2)$\,\,\,\, $1\times L= (1,\overline{2})$ \,\,\,\\
\\ $1\times H_u =  (1,\overline{2})$\,\,\,\, $1\times H_d =  (1,\overline{2})$\,\,\,\,.
\item[(3)] $1\times Q_l=(3, \overline 2)$ \,\,\,\,$2 \times q_l=(3,2)$ \,\,\,\,
$2 \times l=(1,2)$\,\,\,\, $1\times L= (1,\overline{2})$ \,\,\,\\
\\ $2\times H_u =  (1,\overline{2})$\,\,\,\, $2\times H_d =  (1,\overline{2})$\,\,\,\,.
\end{itemize}
Subsequently, we refer to these three setups as the \emph{square quiver},
\emph{triangle quiver} and \emph{triangle quiver with doubled Higgs
sector}, respectively \cite{Ibanez:2001nd,Cremades:2002cs}.

Note, that the second setup, the triangle quiver, does not fulfill
the constraint, equal number of fundamentals and anti-fundamentals
for the $SU(2)$. Thus in this case additional exotics, charged under
the MSSM gauge groups, are required to ensure global consistency. In
the third setup we allow for an additional Higgs pair. Later we will
see that the second Higgs pair is crucial to overcome a
phenomenological problem encountered in the triangle quiver with
just one Higgs pair.

All three setups are analyzed with respect to their Yukawa
couplings. We will see, that various MSSM couplings are absent due
to violation of global $U(1)$ selection rules. For these
perturbatively forbidden couplings we discuss under what
circumstances they can be generated non-perturbatively.
Specifically, we present necessary conditions on the instanton zero
mode structure and analyze their phenomenology with respect to the
experimentally observed hierarchies.

For the triangle quivers, setups (2) and (3), we give local
realizations. As background we choose the orientifold $T^6/{\mathbb
Z}_2 \times {\mathbb Z}_2'$ with Hodge numbers
$(h_{11},h_{12})=(3,51)$, often called the $T^6/{\mathbb Z}_2 \times
{\mathbb Z}_2'$ with discrete torsion, which gives rise to rigid
cycles \cite{Blumenhagen:2005tn}. A brief introduction, which covers
all necessary ingredients for the construction of the local
realizations, is given in appendix \ref{appendix orientifold}. We
show that these local realizations exhibit instantons which carry
the right zero mode structure to induce the perturbatively
forbidden, but desired couplings.

\subsection{Square quiver \label{sqare quiver}}  Such a setup can be realized by the
intersection numbers
\cite{Ibanez:2001nd,Cremades:2002cs}\footnote{Positive intersection
number $I_{ab}=\pi_a \circ \pi_b$ corresponds to a chiral superfield
transforming as $(a,\overline{b})$ under the gauge groups $U(N_a)$
and $U(N_b).$ }
\begin{align*}
& I_{ab}=1 \qquad I_{ab'}=2 \qquad I_{ac}=-3 \qquad I_{ac'}=-3
\qquad I_{bd}=0 \qquad I_{bd'}=-3\\  & \qquad I_{bc}=0 \qquad
I_{bc'}=0 \qquad I^{{\cal N}=2}_{bc'}=1\qquad I_{cd}=-3 \qquad
I_{cd'}=3  \ .
\end{align*}
Note that these intersection numbers correspond to a possible
SUSY extension of the class of non-SUSY models constructed
in \cite{Ibanez:2001nd}. They
 give rise to the spectrum displayed in table \ref{spectrum3}.
\begin{table}[h]
\centering
\begin{tabular}{|c|c|c|c|}
\hline
 sector & matter fields &  transformation & multiplicity\\
\hline \hline
 $ab$                            & $Q_L$  & $(a,\overline{b})$ &$1$ \\
\hline
$ab'$& $q_L$ & $ (a,b) $  &$2$\\
\hline
 $ac$                            & $u_R$  & $(\overline{a},c)$ &$3$  \\
\hline
$ac'$ & $d_R$ & $(\overline{a},{\overline c})$ &$3$ \\
\hline
$bd'$                            & $l$  & $(\overline{b},\overline{d})$ &$3$ \\
\hline
$bc'$                            & $H_u+ H_d$  & $(\overline{b},\overline{c})+(b,c)$ &$1$ \\
\hline
$cd$                            & $E_R$  & $(\overline{c},d)$  &$3$\\
\hline
$cd'$                            & $N_R$  & $(c,d)$  & $3$\\
\hline
\end{tabular}
\caption{Spectrum for the square quiver.} 
\label{spectrum3}
\end{table}\vspace{5pt}

Generically, the anomalous $U(1)$'s acquire a mass via the
Green-Schwarz mechanism and survive as global symmetries. To achieve
the presence of the standard model gauge groups, we require that the
combination
\begin{align}
U(1)_Y= \frac{1}{3}\, U(1)_a -U(1)_c +U(1)_d,
\end{align}
which corresponds to the hypercharge, remains massless. For
simplicity we assume, that this is the only massless $U(1)$. Thus
the gauge symmetry in four dimensional space-time is
\begin{align*}
SU(3)_a \times SU(2)_b \times U(1)_Y\,\,.
\end{align*}
The perturbatively realized couplings are
\begin{align} \nonumber
&<{q^I_L}_{(1,0,0)} \, {H_{u}}_{(-1,-1,0)} \, {u^J_R}_{(0,1,0)} >
\qquad
<{Q_L}_{(-1,0,0)} \, {H_{d}}_{(1,1,0)} \, {d^I_R}_{(0,-1,0)} >\\
&\qquad \qquad \qquad <l^I_{(-1,0,-1)} \, {H_{d}}_{(1,1,0)} \,
{E^J_R}_{(0,-1,1)}> \ .
\end{align}
Note that for generic entries in the corresponding mass matrices at this
level at least the $u,d$ and $s$ quarks remain massless. There are also no
Dirac neutrino masses at this level.
On the other hand stringy instantons could give rise to the
couplings
\begin{align} \nonumber
&<{Q_L}_{(-1,0,0)} \, {H_{u}}_{(-1,-1,0)} \, {u^I_R}_{(0,1,0)} >
\qquad
<{q^I_L}_{(1,0,0)} \, {H_{d}}_{(1,1,0)} \, {d^J_R}_{(0,-1,0)} >\\
&\qquad \qquad \qquad <l^I_{(-1,0,-1)} \, {H_{u}}_{(-1,-1,0)} \,
{N^J_R}_{(0,1,1)}>\,\,
\end{align}
all of which violate the $U(1)_b$ symmetry and hence are perturbatively
absent.  Here the capital letter $I$, $J$ denote the family index and
the subscript indicates the charge under the global $U(1)_b$,
$U(1)_c$ and $U(1)_d$. All three forbidden couplings can be
generated non-perturbatively by three different rigid $O(1)$
instantons $E2_1$, $E2_2$ and $E2_3$. In order to have the right
charged zero mode structure they have to satisfy
\begin{align*}
Q_L \, H_{u} {u^I_R}: &\,\,\,\, \bullet \,\,I_{E2_1b}=-1 \qquad
I_{E2_1a}=I_{E2_1c}=I_{E2_1d}=0\\
& \,\,\,\,\bullet\,\, I_{E2_1b}=-1 \qquad I_{E2'_1a}=I_{E2'_1c}=I_{E2'_1d}=0 \qquad I^{N=2}_{E2'_1c}=1\\
q^I_L \, H_{d} \, {d^J_R}: &\,\,\,\, \bullet\,\,
I_{E2_2b}=1 \qquad I_{E2_2a}=I_{E2_2c}=I_{E2_2d}=0 \\
&\,\,\,\, \bullet\,\,
I_{E2'_2b}=1 \qquad I_{E2'_2a}=I_{E2'_2c}=I_{E2'_2d}=0\qquad I^{N=2}_{E2'_2c}=1\\
l^I \, {H_{u}}\, {N^J_R}: &\,\,\,\, \bullet \,\, I_{E2_3b}=-1 \qquad
I_{E2_3a}=I_{E2_3c}=I_{E2_3d}=0\\
&\,\,\,\, \bullet \,\, I_{E2'_3b}=-1 \qquad
I_{E2'_3a}=I_{E2'_3c}=I_{E2'_3d}=0 \qquad I^{N=2}_{E2'_3d}=1\\
&\,\,\,\, \bullet \,\, I_{E2''_3b}=-1 \qquad
I_{E2''_3a}=I_{E2''_3c}=I_{E2''_3d}=0 \qquad I^{N=2}_{E2''_3c}=1
\\&\,\,\,\, \bullet \,\,
I_{E2'''_3b}=-1 \qquad I_{E2'''_3a}=I_{E2'''_3c}=I_{E2'''_3d}=0
\qquad I^{N=2}_{E2'''_3c}=I^{N=2}_{E2'''_3d}=1\,\,.
\end{align*}

\begin{figure}[h]
\begin{center}
 \includegraphics[width=0.9\textwidth]{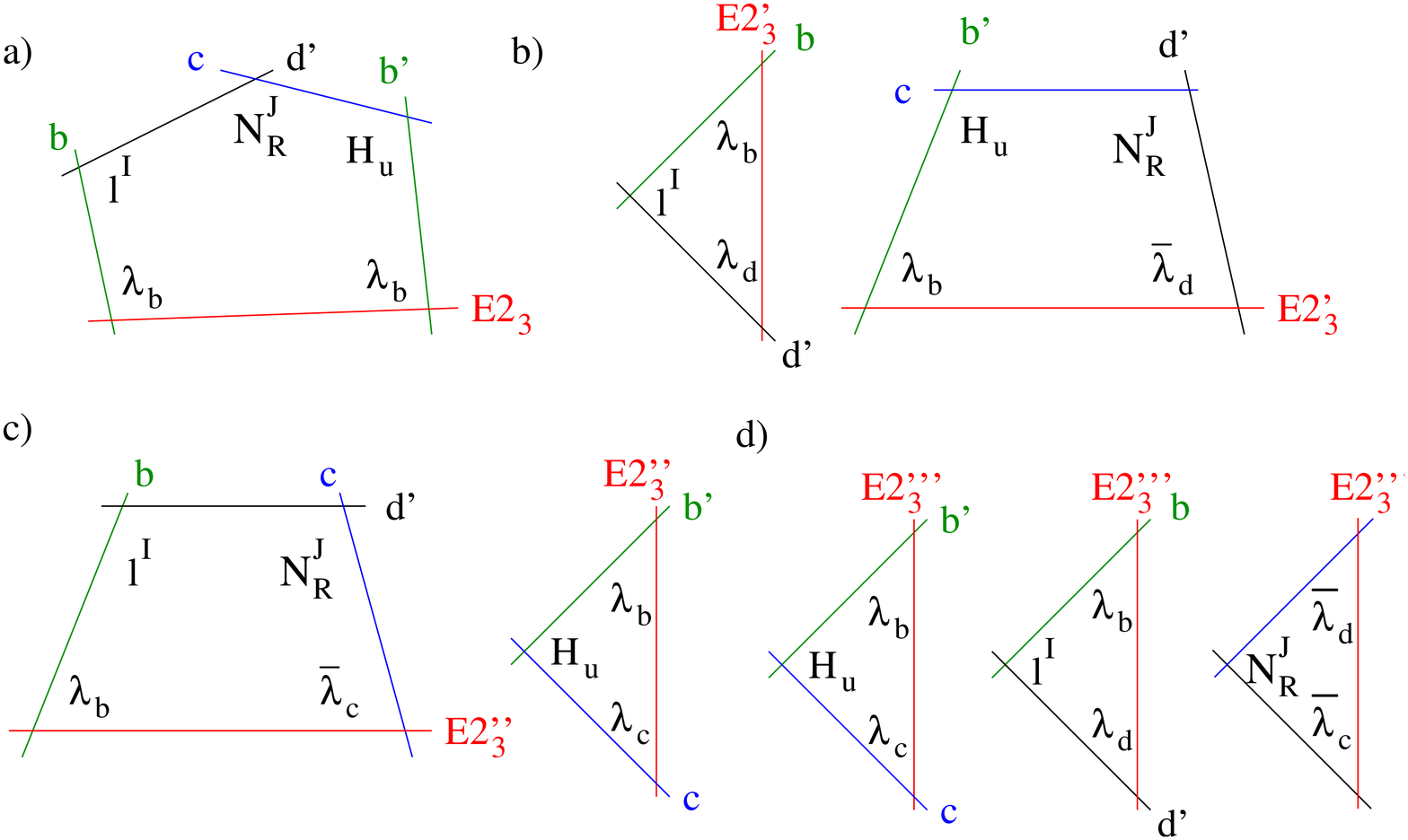}
\end{center}
\caption{\small Instanton induced of the Yukawa coupling $l^I\, H_u
\, N^J_R$ for the square quiver.}\label{leptonsetup1}
\end{figure}

We begin by discussing the leptonic Yukawa coupling $l^I H_d N^J_R$
giving Dirac masses to neutrinos,
which can be generated via the four different instantons listed
above. Figures \ref{leptonsetup1}a-\ref{leptonsetup1}d display for
all four instantons the disk diagrams necessary to soak up all the
charged zero modes. The non-perturbative generation of the Yukawa
couplings $l^I H_d N^J_r$ provides an intriguing explanation for the
smallness of the Dirac Neutrino masses \cite{Cvetic:2008hi}. For
concreteness let us perform the computation of the non-perturbative
contribution arising from an instanton $E2_3$, exhibiting no
vector-like charged zero modes. The measure of the path-integral
contains all instanton zero modes, the four bosonic and two
fermionic universal zero modes, $x^{\mu}$ and $\theta^{\alpha}$, as
well as the two charged fermionic zero modes $\lambda_b$
\begin{align}
\int d^4x \,d^2\theta \, d^2 \lambda_b  \,
 e^{-S^{cl}_{E2_3}}
  \epsilon_{ij}\, \epsilon_{kl}\,<
\lambda^i_b   l^j   N_R H^k_u  \lambda^l_b> e^{Z'}
\end{align}
Here $i$, $j$, $k$ and $l$ denote the gauge indices of the $SU(2)$,
$e^{-S^{cl}_{E2_3}}$ is the suppression factor of the instanton
$E2_3$ and $e^{Z'}$ is the regularized one loop amplitude which can
be interpreted as one-loop Pfaffian. To avoid an overload in
notation we omit the family index\footnote{The reader should keep in
mind that the instanton induces a $3\times 3$ Yukawa coupling
matrix.}. Performing the integral over the charged zero modes
results into
\begin{align}
  \int d^4x \,d^2\theta \,\, Y_{lH_uN_R} \, \epsilon_{il}\,  l^i   H^l_u
  N_R\,\,.
\end{align}
Here $Y_{lH_uN_R}$ contains, apart from the classical suppression
factor and $e^{Z'}$, also the contribution arising from the disk
amplitude $< \lambda_b   l   N_R H_u  \lambda_b>$, which depends via
world-sheet instantons on the open string moduli. To get the desired
small masses for the Neutrinos the instanton suppression factor
should be in the range of $10^{-13}$ to $10^{-11}$. An analogous
analysis can be performed for the other three instantons with the
difference that the charged zero modes get soaked up by more than
one disk diagram. Note, that for the two  instantons $E2'_3$
and $E2'''_3$ the disk diagrams do not contain both matter fields
$l^I$ and $N^J_R$, simultaneously. Thus the induced $3 \times 3$
Yukawa coupling matrix factorizes
\begin{align} Y^{IJ}_{l H_d N_r} =Y^{I} \, Y^{J}
\end{align}
and in the absence of any other instanton the Yukawa coupling $l^I
H_d N^J_r$ gets generated only for one family. Later we will see,
that the factorization of instanton induced Yukawa coupling
potentially gives a natural explanation for hierarchies within the
class of non-perturbative Yukawa couplings.
\begin{figure}[h]
\begin{center}
 \includegraphics[width=0.9\textwidth]{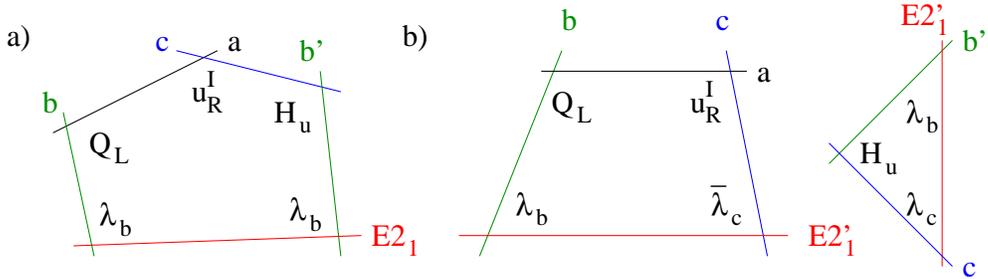}
\end{center}
\caption{\small Instanton induced Yukawa coupling $Q_L\, H_u \, u_R$
for the square quiver.}\label{fourstacksetupone}
\end{figure}

Let us turn to the quark Yukawa couplings $Q_L\,H_{u}\,u^I_R$ and
${q^I_L}\,{H_{d}}\,{d^J_R}$. The charged zero modes of $E2_1$ get
soaked up by one disk diagram depicted in figure
\ref{fourstacksetupone}a inducing the coupling $Q_L\,H_{u}\,u^I_R$.
For $E2'_1$, exhibiting  two additional vector-like zero modes
$\lambda_c$ and ${\overline \lambda}_c $, the instanton zero modes
get saturated via two disk diagrams (see figure
\ref{fourstacksetupone}b). Analogously, the other missing quark
Yukawa coupling, ${q^I_L}\,{H_{d}}\,{d^I_R}$ is generated by $E2_2$
or $E2'_2$ respectively. In case both instantons $E2_i$ and $E2'_i$
are present, the one with the smaller suppression factor
$e^{S^{cl}_E2}$ gives the dominant contribution.


Let us now  discuss the hierarchies of the quark Yukawa couplings.
The fact that only one D-quark gets perturbatively a mass suggests
to identify $Q_L$ with the left-handed quarks of the third
generation. So  $q^I_L$'s should correspond to the two lightest
generations. Then, the suppression factor $e^{-S^{cl}_{E2_1}}$
($e^{-S^{cl}_{E2'_1}}$) gives a natural explanation for the
smallness of the Yukawa couplings ${q^I_L}\, H_{d} \, {d_R}$ for the
two lightest  D-quarks compared to the heaviest. However, due to the
same reasoning one would expect a smaller Yukawa coupling
$Q_L\,{H_{u}}\,u^I_R$ for the heaviest family. Thus, in
contradiction to observations, a smaller mass for the top-quark,
than for the up- and charm-quark gets generated (or a very large
mixing for the third generation is obtained). To be more concrete
the perturbatively generated Yukawa coupling matrices take the form
\begin{align}
Y^P_{u^I_L H_u u^J_R}=\left(\begin{array}{ccc} A^u_{11}&A^u_{12}&
A^u_{13}\\
A^u_{21}&A^u_{22}& A^u_{23}\\
0&0&0\\
\end{array}\right)
\qquad Y^{P}_{d^I_L H_d d^J_R}=\left(\begin{array}{ccc}
0&0&0\\
0&0&0\\
A^d_{31}&A^d_{32}& A^d_{33}\\
\end{array}\right)
\end{align}
while the non-perturbatively are given by
\begin{align}
Y^{NP}_{u^I_L H_u u^J_R}= \left(\begin{array}{ccc}
0&0&0\\
0&0&0\\
B^u_{31}&B^u_{32}& B^u_{33}\\
\end{array}\right)
\qquad Y^{NP}_{d^I_L H_d d^J_R}=\left(\begin{array}{ccc}
B^d_{11}&B^d_{12}&
B^d_{13}\\
B^d_{21}&B^d_{22}& B^d_{23}\\
0&0&0\\
\end{array}\right).
\end{align}
The total $3 \times 3$ Yukawa coupling matrices are given by the sum
of the respective perturbative and non-perturbative part.
Generically, due to the instanton suppression factor we expect
$A_{ij}\gg B_{kl}$ for arbitrary $i,j,k$ and $l$. Determining the
eigenvalues and eigenstates of this matrix reveals a large mixing
between the first and third generation which is not observed in
nature. Large world-sheet instanton suppressions for specific
perturbative realized couplings might relax the expectation
$A_{ij}\gg B_{kl}$ and surmount the encountered problem of a too
large family mixing. Nevertheless to achieve agreement with
experimental observations a lot of fine-tuning is required.


\subsection{Triangle quiver\label{setup 2}}
Let us turn the triangle quiver with exactly one Higgs pair, where
we overcome the issue, of a large fine-tuning to match experimental
observations, encountered in the previous setup. Consider the
intersection numbers \cite{Cremades:2002cs}
\begin{align} \nonumber
&I_{ab}=1 \qquad I_{ab'}=2 \qquad I_{ac}=-3 \qquad I_{ac'}=-3 \qquad
I_{bd}=-1 \\  \label{intersection number higgs1} & I_{bd'}=2 \qquad
I_{bc}=-1 \qquad I_{bc'}=-1\qquad I_{cd}=3 \qquad I_{cd'}=-3\,\,,
\end{align}
which give rise to the spectrum displayed in table \ref{spectrum
ibanezmodel}. Tentatively we can asign
$Q_L$ to  the left-handed quarks of the
lightest family, while the $q_l$'s would be  the left-handed quarks of the
other two families. The $L$ is the lepton-doublet of the heaviest
family, and the $l$ denotes the $SU(2)$ doublet of the two lightest
generations. Note that global consistency requires the existence of
additional exotics, charged under the $SU(2)$.\begin{table}[h]
\centering
\begin{tabular}{|c|c|c|c|}
\hline
 sector & matter fields &  transformation & multiplicities \\
\hline \hline
 $ab$                            & $Q_L$  & $(a,\overline{b})$ &$1$ \\
\hline
 $ab'$                            & $q_L$ &  $(a,b)$ & $2$ \\
\hline
 $ac$                            & $u_R$  &$(\overline{a},c)$ & $3$  \\
\hline
$ac'$                            & $d_R$  & $(\overline{a},\overline{c})$ & $3$ \\
\hline
$bd$                            & $L$  & $(\overline{b},d)$  & $1$\\
\hline
$bd'$                            & $l$  & $(b,d)$  & $2$\\
\hline
$bc$                            & $H_d$  & $(\overline{b},c)$ & $1$ \\
\hline
$bc'$                            & $H_u$  & $(\overline{b},\overline{c})$ & $1$ \\
\hline
$cd$                            & $N_R$  & $(c,\overline{d})$  & $3$ \\
\hline
$cd'$                            & $E_R$  & $(\overline{c},\overline{d})$ & $3$  \\
\hline
\end{tabular}
\caption{Spectrum for the triangle quiver.} 
\label{spectrum ibanezmodel}
\end{table}\vspace{5pt}
The hypercharge is given by the combination
\begin{align}
U(1)_Y=\frac{1}{3}\,U(1)_a -U(1)_c-U(1)_d\,\,.
\end{align}
Assuming, that all  other extra  $U(1)$'s become massive via
couplings to RR-fields,  the gauge symmetry in four-dimensional
spacetime is
\begin{align}
SU(3)_a \times SU(2)_b \times U(1)_Y\,\,.
\end{align}

The perturbative Yukawa couplings are
\begin{align} \nonumber
&<{q_L}_{(1,0,0)} \, {H_{u}}_{(-1,-1,0)} \, {u_R}_{(0,1,0)} > \qquad
<{q_L}_{(1,0,0)} \, {H_{d}}_{(-1,1,0)} \, {d_R}_{(0,-1,0)} > \\
 &  <l_{(1,0,1)} \, {H_{d}}_{(-1,1,0)} \,
{E_R}_{(0,-1,-1)}> \qquad <l_{(1,0,1)} \, {H_{u}}_{(-1,-1,0)} \,
{N_R}_{(0,1,-1)}>, \label{perturbative realized}
\end{align}
where the subscripts denote as before the charge under the global
$U(1)_b$, $U(1)_c$ and $U(1)_d$.
Note that all fermion mass matrices have then the general structure
\beq
m_{U,D,L,N}\ \simeq \
\left(
\begin{array}{ccc}
A_{11} &  A_{12} &  A_{13} \\
A_{21} & A_{22} & A_{23} \\
0 & 0 & 0
\end{array}
\right)
\label{yukper}
\eeq
at the perturbative level.
The
couplings which are
 perturbatively forbidden (since they violate $U(1)_b$)
are
\begin{align} \nonumber
&<{Q_L}_{(-1,0,0)} \, {H_{u}}_{(-1,-1,0)} \, {u_R}_{(0,1,0)}  >
\qquad
<{Q_L}_{(-1,0,0)} \,  {H_{d}}_{(-1,1,0)} \, {d_R}_{(0,-1,0)}  > \\
\label{perturbative forbidden}& <L_{(-1,0,1)} \, {H_{d}}_{(-1,1,0)}
\, {E_R}_{(0,-1,-1)}> \qquad <L_{(-1,0,1)} \, {H_{u}}_{(-1,-1,0)} \,
{N_R}_{(0,1,-1)}>
\end{align}
and fill the remaining entries in eq.(\ref{yukper}).

Such a setup can be locally realized on the $T^6/{\mathbb Z}_2
\times {\mathbb Z}_2'$ orientifold where the first two tori are
untilted, the third one is tilted and the orientifold charges are
chosen to be
\begin{align}
\eta_{\Omega {\cal R}}=1 \qquad  \eta_{\Omega {\cal R} \theta}=1
\qquad  \eta_{\Omega {\cal R} \theta'}=-1 \qquad \eta_{\Omega {\cal
R} \theta\theta'}=1 \label{orientifold charge setup 1}\,\,,
\end{align}
satisfying \eqref{etaconst}. In appendix \ref{appendix orientifold}
we present a brief review of this orbifold, which includes a brief
discussion of rigid cycles, details of the orientifold action as
well as the computation of chiral intersection numbers.

The spacetime-filling D6-branes are given by fractional branes,
branes that carry charge only under one twisted sector. These
fractional branes wrap the bulk cycles, displayed in table
\ref{wrapping numbers for ibanezmodel}
\begin{table}[h] \centering
\begin{tabular}{|c|c|c|c|}
\hline
brane & $(n_1,m_1)$ & $(n_2,m_2)$&  $(n_3,\tilde{m}_3)$\\
\hline \hline
 $N_a=3$                            & $(1,0)$  &$(3,1)$& $(3,-1/2)$  \\
\hline
 $N_b=2$                            & $(1,1)$ & $(1,0)$& $(1,-1/2)$  \\
\hline
 $N_c=1$                            & $(0,1)$  & $(0,-1)$& $(2,0)$  \\
\hline
 $N_d=1$                            & $(1,0)$  &$(3,1)$& $(3,-1/2)$  \\
\hline
\end{tabular}
\caption{Bulk wrapping numbers} 
\label{wrapping numbers for ibanezmodel}
\end{table}\vspace{5pt}
and their complete homology classes are given by
\begin{align} \nonumber
\pi^F_a&= \frac{1}{2}\, [(1,0),(3,1),(3,-1/2)] +\frac{1}{2} \left(
\sum_{i,j \in(2,4)\times (1,4)}\Big[
\alpha^{\theta}_{ij}\times[(3,-1/2)\Big]\right)
\\ \nonumber
\pi^F_b&= \frac{1}{2} \,[(1,1),(1,0),(1,-1/2)] +\frac{1}{2}\left(
\sum_{i,j \in(1,4)\times (1,3)}\Big[
\alpha^{\theta\theta'}_{ij}\times[(1,0)\Big]\right)
\\ \label{configuration 1}
\pi^F_c&= \frac{1}{2} \,[(0,1),(0,-1),(2,0)]
\,\,\,\,\,\,+\frac{1}{2} \left(\sum_{i,j \in(1,2)\times (3,4)}
\Big[\alpha^{\theta'}_{ij}\times[(0,1)\Big]\right)
\\ \nonumber
\pi^F_d &= \frac{1}{2} \,[(1,0),(3,1),(3,-1/2)] +\frac{1}{2}\left(
\sum_{i,j \in(2,4)\times (2,3)}
\Big[\alpha^{\theta}_{ij}\times(3,-1/2)\Big]\right)\,\,.
\end{align}
The configuration gives rise to a $U(3)_a \times U(2)_b \times
U(1)_c \times U(1)_d$ gauge theory in four-dimensional spacetime,
with the spectrum, computed using equation \eqref{intersection
fractional brane} in appendix \ref{appendix orientifold}, displayed
in table \ref{spectrum ibanezmodel}.

Global consistency requires the presence of additional branes. For
the hypercharge to be realized as a local symmetry the combination
\begin{align}
U_Y(1)=\frac{1}{3}\, U(1)_a -  U(1)_c -U(1)_d + \sum_{x} \,c_x
\,U(1)_x
\end{align}
has to remain massless. The $U(1)_x$ denote the anomalous $U(1)$'s
arising from the additional branes, required for tadpole
cancellation. Again for simplicity we assume that this is the only
massless combination, all other $U(1)$'s become massive and survive
only as global symmetry. With the choice
\begin{align}
U_1=\frac{u}{2} \qquad U_2=\frac{u}{2} \qquad U_3=u
\end{align}
for the complex structure moduli all $D6$-branes are aligned to the
orientifold plane, thus ${\cal N}=1 $ supersymmetry in 4D is
ensured.

In this local D6-brane configuration the couplings, listed in
\eqref{perturbative realized}, are perturbatively realized, while
the Yukawa couplings \eqref{perturbative forbidden} violate global
$U(1)$ selection rules, thus are perturbatively absent. In order to
match experimental observation we expect them to be generated via
non-perturbative effects.

Before investigating the details of this local realization, let us
analyze the non-perturbative effects for a generic triangle quiver,
realizing the intersection pattern \eqref{intersection number
higgs1}. There are potentially two different types of instanton,
$E2_1$ and $E2_2$, both generating all the missing couplings in
\eqref{perturbative forbidden}. Their intersection pattern with the
matter $D6$-branes has to satisfy
\begin{align}
&I_{E2_1b}=-1 \qquad I_{E2_1a}=I_{E2_1c}=I_{E2_1d}=0 \label{instanton 1}\\
&I_{E2_2b}=-1 \qquad I_{E2_2a}=I_{E2_2c}=I_{E2_2d}=0 \qquad I^{{\cal
N}=2}_{E2_1c}=1 \label{instanton 2}\,\,.
\end{align}
In addition there are two classes of instantons $E2_3$ and $E2_4$
with the intersection pattern
\begin{align}
&I_{E2_3b}=-1 \qquad I_{E2_3a}=I_{E2_3c}=I_{E2_3d}=0 \qquad I^{{\cal
N}=2}_{E2_3d}=1 \label{instanton 3} \\&I_{E2_4b}=-1 \qquad
I_{E2_4a}=I_{E2_4c}=I_{E2_4d}=0 \qquad I^{{\cal
N}=2}_{E2_4c}=I^{{\cal N}=2}_{E2_4d}=1 \label{instanton 4}
\end{align}
which generate only the lepton Yukawa coupling $L\, {H_{d}}\,{E_R}$
and $L\,H_{u}\,{N_R}$.

\begin{figure}[h]
\begin{center}
 \includegraphics[width=0.9\textwidth]{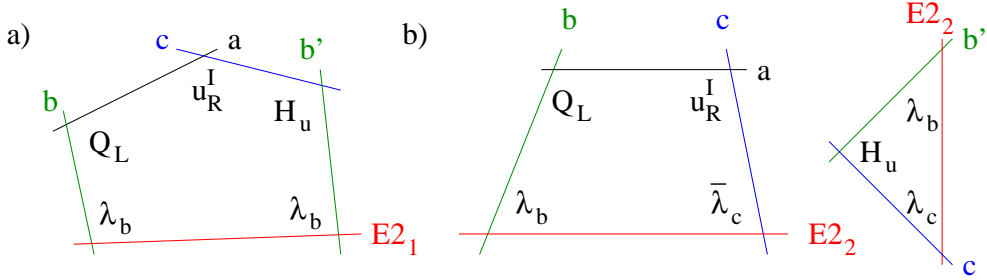}
\end{center}
\caption{\small Instanton induced Yukawa coupling $Q_L\, H_u \,
u^I_R$ for the triangle quiver.}\label{quarksetup2}
\end{figure}

Let us now analyze the generation of the U-quark Yukawa coupling
$Q_L\, H_u \, u^I_R$ in more detail. An analogous discussion applies
to the coupling $Q_L\, H_d \, d^I_R$. For the instanton $E2_1$
satisfying \eqref{instanton 1} the two charged zero modes
$\lambda_b$ get soaked up by one disk diagram displayed in figure
\ref{quarksetup2}a. The path integral takes the form
\begin{align}
\int d^4x \,d^2\theta \, d^2 \lambda_b  \,
 e^{-S^{cl}_{E2_1}}
  \epsilon_{ij}\, \epsilon_{kl}\,<
\lambda^i_b \, Q^j_L\, u^I_R \,H^k_u \lambda^l_b> e^{Z'}
\end{align}
 which, after performing the integral over the charged fermionic zero modes, results into
\begin{align}
 \int d^4x \,d^2\theta \, \, Y_{Q_L H_u u^I_R} \,\epsilon_{il} \,
Q^i_L\, H^l_u \, u^I_R \,\,.
\end{align}
The indices $i$, $j$, $k$ and $l$ label again the flavor charge, the
capital letter denotes the family and $Y_{Q_L\, H_u \, u^I_R}$ is
the final effective Yukawa coupling, including the suppression
factor $e^{-S^{cl}_{E2_1}}$, the loop contribution $e^{Z'}$ as well
as the world-sheet contribution arising from the disk diagram $<
\lambda_b \, Q_L\, u^I_R \,H_u \lambda_b>$.

For the instanton $E2_2$, with an additional pair of vector-like
charged zero modes, one performs a similar computation with the
difference that the four charged zero modes get saturated via two
disk diagrams (see figure \ref{quarksetup2}b)
\begin{align}
\int d^4x \,d^2\theta \, d^2 \lambda_b  \, d\overline{\lambda}_c \,
d\lambda_c\, e^{-S^{cl}_{E2_2}}
  \epsilon_{ij}\,<
\lambda^i_b \, Q^j_L\, u^I_R \overline{\lambda}_{c} > \,
\epsilon_{kl}\, <\lambda^k_b  \,H^l_u \lambda_c> e^{Z'}\,\,,
\end{align}
which gives the superpotential contribution
\begin{align}
 \int d^4x \,d^2\theta \, \, Y'_{Q_L H_u  u^I_R} \,\epsilon_{il} \,
Q^i_L\, H^l_u \, u^I_R \,\,.
\end{align}
In case, both types of instantons are present, the one, which wraps
the smaller cycle in the internal manifold, thus exhibiting a
smaller suppression factor in the path integral, gives the dominant
contribution to the Yukawa couplings. To match experimental data the
suppression factor of the dominant contribution is expected to be of
the order $10^{-2}-10^{-5}$.

An analogous analysis applies to the lepton Yukawa couplings  $l^I
\, {H_d} \, {E_R}^J$ and ${l}^I \, {H_u} \, {N_R}^J$, with the
difference that there are potentially two additional classes of
instantons $E2_3$ and $E2_4$ which can induce these couplings. Again
the dominant contribution arises from the instanton, which wraps the
smallest cycle in the internal manifold. Note, that in opposite to
the previous setup here the Dirac Neutrino masses are expected to be
of the same order as the masses of charged  lepton.
 One possibility to obtain small masses for neutrinos is due
to the so called see-saw mechanism. For this mechanism to work a
large Majorana mass term for the right-handed Neutrinos is required.
Such a mass term could be generated via an instanton $E2_x$ with the
intersection pattern \cite{Blumenhagen:2006xt,Ibanez:2006da}
\begin{align}
I_{E2_xc}=I_{E2_{x}d}=2 \qquad \,I_{E2_xa}=I_{E2_{x}b}=0.
\label{majorana mass}
\end{align}
With a suppression factor of the order $10^{-8}$ to $10^{-5}$ the
Majorana mass term $M_{N_R}$ for the right-handed Neutrino lies in
the range $10^{10}GeV< M_{N_R}< 10^{13}GeV$. Together with the Dirac
masses for the Neutrinos, which are of the electroweak scale
$(0.01-1)GeV$ that would give  the see-saw Neutrino masses in the
desired range $(10^{-2}-10^{-1})eV$.

After the analysis for a generic triangle quiver satisfying the
intersection pattern \eqref{intersection number higgs1}, let us turn
to our concrete realization and examine if it exhibits instantons,
inducing the perturbatively forbidden couplings. Indeed, as we will
see momentarily the local realization comprises one representant for
each of the first two classes, $E2_1$ and $E2_2$.

An instanton wrapping the cycle
\begin{align}\label{instanton 1 setup 1}
\pi_{E2_1} =&\,\frac{1}{4} \,[(1,0)(0,1)(0,-1)] + \frac{1}{4}\left(
\sum_{i,j \in (1,3)\times(3,4)} \Big[\alpha^{\theta}_{ij} \times
(0,-1)\Big]\right) \\ \nonumber +&\, \frac{1}{4} \left(\sum_{i,j \in
(3,4)\times(1,2)}\Big[ \alpha^{\theta'}_{ij} \times
(1,0)\Big]\right) + \frac{1}{4}\left( \sum_{i,j \in
(1,3)\times(1,2)} \Big[ \alpha^{\theta\theta'}_{ij} \times [(0,1)
\Big]\right)
\end{align}
gives rise to the intersection pattern in \eqref{instanton 1}. Note,
that all ${\cal N}=2$ modes between the instanton and the
$D6$-branes are massive (see also appendix \ref{appendix N=2}).
Moreover, the cycle $\pi_{E2_1}$ is rigid and invariant under the
orientifold action $\Omega {\cal R}$, thus the instanton contains
only the four bosonic zero modes $x^{\mu}$ and the two fermionic
$\theta^{\alpha}$. Therefore, the instanton $E2_1$ indeed gives rise
to all the perturbatively missing couplings in \eqref{perturbative
forbidden}. Its suppression factor is
\begin{align}
e^{-S^{cl}_{E2_1}}=e^{-\frac{2 \pi}{l^3_s\, g_s} \,
\text{Vol}_{E2_1}} =e^{-\frac{2\pi}{\alpha_a}\,
\frac{\text{Vol}_{E2_1}}{\text{Vol}_{D6_a}}}\,\,,
\label{suppression1}
\end{align}
where the ratio $\frac{\text{Vol}_{E2_1}}{\text{Vol}_{D6_a}}$ is
given by
\begin{align}
\frac{\text{Vol}_{E2_1}}{\text{Vol}_{D6_a}}=
\frac{1}{2}\left(\prod_I \left[ \frac{(n^I_{E2_1})^2 + (\widetilde
m^I_{E2_1})^2 U_I^2}{(n^I_{a})^2 + (\widetilde m^I_{a})^2
U_I^2}\right]\right) ^{1/2}\,\,. \label{supprssion 2}
\end{align}
The factor $\frac{1}{2}$ is due to the fact that the $D6$-brane
wraps a fractional cycle while the instanton a rigid one (see the
different prefactors in equation \eqref{configuration 1} and
\eqref{instanton 1 setup 1}). If we assume $\alpha_a=1/24$ at string
scale and the complex structure moduli $U_3=u$ stabilized in the
range $(1-1.5)$ we get the desired suppression factor
\begin{align}
e^{-S^{cl}_{E2_1}} \sim \, 10^{-2}- 10^{-5}.
\end{align}

There is also a representant $E2_2$ for the second class of
instantons. It wraps the cycle
\begin{align}\label{instanton 2 setup 1}
\pi_{E2_2} =&\,\frac{1}{4} \,[(1,0)(0,1)(0,-1)] + \frac{1}{4}\left(
\sum_{i,j \in (1,3)\times(1,2)} \Big[\alpha^{\theta}_{ij} \times
(0,-1)\Big]\right) \\ \nonumber +&\, \frac{1}{4} \left(\sum_{i,j \in
(1,2)\times(1,2)}\Big[ \alpha^{\theta'}_{ij} \times
(1,0)\Big]\right) + \frac{1}{4}\left( \sum_{i,j \in
(1,3)\times(1,2)} \Big[ \alpha^{\theta\theta'}_{ij} \times [(0,1)
\Big]\right)
\end{align}
and gives rise to the same chiral intersection pattern as the
instanton $E2_1$, but there also vectorlike pairs of charged zero
modes in the ${\cal N}=2$ sector between instanton $E2_2$ and the
$D6$-branes. We relegate the detailed analysis of the ${\cal N}=2$
instanton zero mode sector to the appendix \ref{appendix N=2}. There
we show that indeed exactly one vectorlike pair $\lambda_c$,
${\overline \lambda}_c$ arises at the intersection $E2_2$ and $c$.
Thus, such an instanton induces as well as $E2_1$ all the missing
couplings in \eqref{perturbative forbidden}. Since the $E2_2$ wraps
the same bulk-cycle as $E2_1$ their suppression factors are equal
\begin{align}
e^{-S^{cl}_{E2_1}}= e^{-S^{cl}_{E2_2}} \sim 10^{-2}- 10^{-5}
\end{align}
and their contributions to the Yukawa couplings are expected of the
same order.

The non-perturbative contributions depend via the disk diagrams on
the world-sheet instantons, thus on the area, that the diagrams
enclose. Therefore the dominant contribution to the Yukawa couplings
arises from that instanton, whose disk-diagrams encloses the
smallest area.


This concrete realization does not exhibit an instanton, which could
give rise to large Majorana mass term for the right-handed Neutrino
$N_R$. Thus at this level the Neutrino masses are of the same order
as the masses for their doublet partner. Note also, that in this
explicit local setup the perturbative contribution to fermion  mass
matrices in eq.(\ref{yukper}) factorizes, due to the fact that the
non-trivial intersections for the family index carrying matter
fields occur in different two tori \cite{Cremades:2003qj}. Thus, the
perturbative matrix has rank one and one fermion per generation
remains massless even after instanton corrections are taken into
account.

\begin{figure}[h]
\begin{center}
 \includegraphics[width=0.9\textwidth]{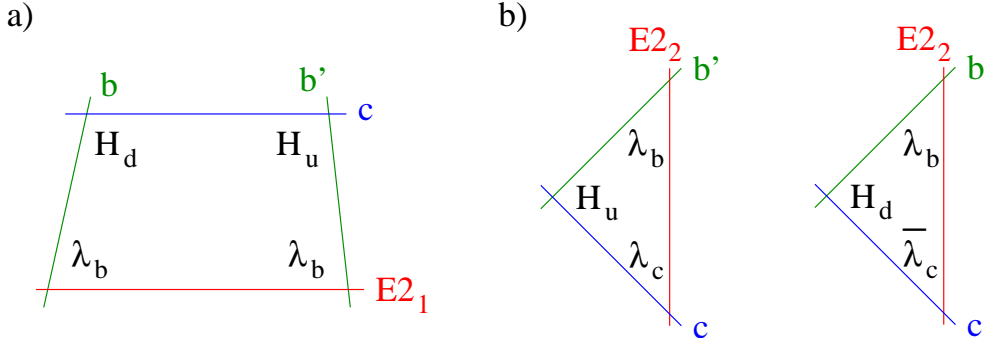}
\end{center}
\caption{\small Non perturbative generation of the $\mu$-term for
the triangle quiver.}\label{Higgssetupone}
\end{figure}

Finally let us  draw attention to a phenomenological drawback of the
simplest triangle quiver. The Yukawa coupling inducing instantons,
$E2_1$ and $E2_2$, generate also a $\mu$-term for the Higgs pair
(see figure \ref{Higgssetupone})
\begin{align}
<{H_u}_{(-1,-1)}\, {H_{d}}_{(-1,1)}>\,\,,
\end{align}
which is far too large. To match the Yukawa couplings with
experimental observations, we expect the instanton suppression
factor $e^{-S^{cl}_{E2}}$ to be in the range of $10^{-2}$ to
$10^{-5}$. But then the same instanton would induce a $\mu$-term of
the order $(10^{13} - 10^{16}) GeV$ rather than $(10^2-10^3)GeV$. In
the next section we examine a setup, where we allow for a second
Higgs pair and show that one can generate all the missing Yukawa
couplings without inducing a too large $\mu$-term for the Higgs
pair.

\subsection{Triangle quiver with doubled Higgs sector\label{setup 3}}

A way to avoid the previously encountered problem of generating a
too large $\mu$-term is to allow for a doubled Higgs sector. The
hope would be, that the Yukawa coupling generating instanton induces
a $\mu$-term matrix which factorizes. Thus it only generates for one
Higgs pair a $\mu$-term of the order $10^{13} - 10^{15} GeV $, while
the other one, the physical one, remains massless. The $\mu$-term
for the physical MSSM-Higgs pair is generated by a different
instanton, which wraps a larger cycle, leading to a larger
suppression factor and possibly to a $\mu$-term of the order
$(10^2-10^3) \, GeV$. If the supermassive unphysical Higgs doublets
do not acquire a vev they can be integrated out and do not affect
the low energy effective action.

In the following we will discuss the realization of such a setup.
The intersection numbers are equivalent to the previous case up to
the $bc$ and $bc'$-sector, which are doubled, respectively,
\begin{align} \nonumber
&I_{ab}=1 \qquad I_{ab'}=2 \qquad I_{ac}=-3 \qquad I_{ac'}=-3 \qquad
I_{bd}=-1 \\  \label{intersection number higgs2} & I_{bd'}=2 \qquad
I_{bc}=-2 \qquad I_{bc'}=-2\qquad I_{cd}=3 \qquad I_{cd'}=-3\,\,.
\end{align}
Note, that this implies the same number of $2$ and $\overline{2}$,
fulfilling the constraint, arising from tadpole cancellation. Table
\ref{spectrum ibanezmodel 2 higgs} displays the origin, the
transformation behavior and the multiplicities of the matter fields.
\begin{table}[h] \centering
\begin{tabular}{|c|c|c|c|}
\hline
 sector & matter fields &  transformation & multiplicities\\
\hline \hline
 $ab$                            & $Q_L$  & $(a,\overline{b})$ & $1$ \\
\hline
 $ab'$                            & $q_L$ & $(a,b)$ & $2$ \\
\hline
 $ac$                            & $u_R$  & $(\overline{a},c)$ & $3$ \\
\hline
$ac'$                            & $d_R$  & $(\overline{a},\overline{c})$ & $3$  \\
\hline
$bd$                            & $L$  & $(\overline{b},d)$ & $1$ \\
\hline
$bd'$                            & $l$  & $(b,d)$  & $2$\\
\hline
$bc$                            & $H_d$  & $(\overline{b},c)$ & $2$ \\
\hline
$bc'$                            & $H_u$  & $(\overline{b},\overline{c})$ & $2$ \\
\hline
$cd$                            & $N_R$  & $(c,\overline{d})$ & $3$  \\
\hline
$cd'$                            & $E_R$  & $(\overline{c},\overline{d})$ & $3$ \\
\hline
\end{tabular}
\caption{Spectrum} 
\label{spectrum ibanezmodel 2 higgs}
\end{table}\vspace{5pt}
Again, we require that apart from the hypercharge
\begin{align}
U(1)_Y= \frac{1}{3} \, U(1)_a -U(1)_c -U(1)_d
\end{align}
all other U(1)'s become massive and survive only as global symmetries.
The perturbatively allowed couplings are
\begin{align} \nonumber
&<{q_L}_{(1,0,0)} \, {H^{I}_{u}}_{(-1,-1,0)} \, {u_R}_{(0,1,0)} >
\qquad <{q_L}_{(1,0,0)} \, {H^{I}_{d}}_{(-1,1,0)} \,
{d_R}_{(0,-1,0)} >\\\label{perturbative allowed two higgs} &
<l_{(1,0,1)} \, {H^I_{d}}_{(-1,1,0)} \, {E_R}_{(0,-1,-1)}> \qquad
<l_{(0,1,0)} \, {H^I_{u}}_{(0,-1,-1)} \, {N_R}_{(0,1,-1)}>\,\,.
\end{align}
The $U(1)$ selection rules violating and thus forbidden couplings
are given by
\begin{align} \nonumber
&<{Q_L}_{(-1,0,0)} \, {H^I_{u}}_{(-1,-1,0)} \, {u_R}_{(0,1,0)} >
\qquad <{Q_L}_{(-1,0,0)} \, {H^I_{d}}_{(-1,1,0)} \, {d_R}_{(0,-1,0)}
> \\\label{perturbative forbidden 2 Higgs}  &
<L_{(-1,0,1)} \, {H^I_{d}}_{(-1,1,0)} \, {E_R}_{(0,-1,-1)}> \qquad
<L_{(-1,0,1)} \, {H^I_{u}}_{(-1,-1,0)} \, {N_R}_{(0,1,-1)}>\,\,,
\end{align}
where $I$ denotes the family index of the Higgs doublets and we
suppress the matter field family index.

For the background $T^6/{\mathbb Z}_2 \times {\mathbb Z}_2'$
orientifold, with the orientifold charges chosen as before (see
equation \eqref{orientifold charge setup 1}) and the last two tori
tilted, a local realization of the intersection numbers
\eqref{intersection number higgs2} is given by four fractional
D6-branes wrapping the bulk-cycles displayed in table \ref{wrapping
numbers for ibanezmodel 2}.

\begin{table}[h] \centering
\begin{tabular}{|c|c|c|c|}
\hline
brane & $(n_1,m_1)$ & $(n_2,m_2)$&  $(n_3,\tilde{m}_3)$\\
\hline \hline
 $N_a=3$                            & $(1,0)$  &$(3,1/2)$& $(3,-1/2)$  \\
\hline
 $N_b=2$                            & $(1,1)$ & $(2,0)$& $(1,-1/2)$  \\
\hline
 $N_c=1$                            & $(0,1)$  & $(0,-1)$& $(2,0)$  \\
\hline
 $N_d=1$                            & $(1,0)$  &$(3,1/2)$& $(3,-1/2)$  \\
\hline
\end{tabular}
\caption{Bulk wrapping numbers} 
\label{wrapping numbers for ibanezmodel 2}
\end{table}\vspace{5pt}
Their complete homology classes are given by
\begin{align} \nonumber
\pi^F_a&= \frac{1}{2}\, [(1,0),(3,1/2),(3,-1/2)] +\frac{1}{2} \left(
\sum_{i,j \in(2,4)\times (1,3)}\Big[
\alpha^{\theta\theta'}_{ij}\times(3,1/2)\Big]\right)
\\ \nonumber
\pi^F_b&= \frac{1}{2} \,[(1,1),(2,0),(1,-1/2)] \,\,\,\,\,\,
+\frac{1}{2}\left( \sum_{i,j \in(1,4)\times (3,4)}\Big[
\alpha^{\theta}_{ij}\times(1,-1/2)\Big]\right)
\\ \label{configuration 2}
\pi^F_c&= \frac{1}{2} \,[(0,1),(0,-1),(2,0)]
\,\,\,\,\,\,\,\,\,\,\,\,+\frac{1}{2} \left(\sum_{i,j \in(1,2)\times
(3,4)} \Big[\alpha^{\theta'}_{ij}\times(0,1)\Big]\right)
\\ \nonumber
\pi^F_d &= \frac{1}{2} \,[(1,0),(3,1/2),(3,-1/2)] +\frac{1}{2}\left(
\sum_{i,j \in(2,4)\times (2,4)}
\Big[\alpha^{\theta\theta'}_{ij}\times(3,1/2)\Big]\right)\,\,
\end{align}
and again we assume that the hypercharge
\begin{align}
U_Y(1)=\frac{1}{3}\, U(1)_a -  \,U(1)_c - \,U(1)_d + \sum_{x} \,c_x
\,U(1)_x
\end{align}
remains massless after including the additional branes required for
global consistency. To ensure supersymmetry we choose the complex
structure moduli to be
\begin{align}
U_1=\frac{u}{2} \qquad U_2=u \qquad U_3=u\,\,.
\end{align}

The perturbatively forbidden couplings in \eqref{perturbative
forbidden 2 Higgs} can be generated by two classes of instantons,
either by an instanton with only two charged zero modes
\begin{align}
I_{E2_2b}=-1 \qquad I_{E2_2a}=I_{E2_2c}=I_{E2_2d}=0
\label{constraints two higgs 1}.
\end{align}
or by an instanton with four charged zero modes
\begin{align}I_{E2_2b}=-1 \qquad
I_{E2_2a}=I_{E2_2c}=I_{E2_2d}=0\qquad I^{{\cal N}=2}_{E2_2c}=1
\label{constraints two higgs 2}\,\,.
\end{align}
As for the triangle quiver with just one Higgs pair there are two
additional classes of instantons
\begin{align}
&I_{E2_3b}=-1 \qquad I_{E2_3a}=I_{E2_3c}=I_{E2_3d}=0 \qquad I^{{\cal
N}=2}_{E2_3d}=1 \label{constraints two higgs 3}\\ &I_{E2_4b}=-1
\qquad I_{E2_4a}=I_{E2_3c}=I_{E2_4d}=0 \qquad I^{{\cal
N}=2}_{E2_4c}=I^{{\cal N}=2}_{E2_4d}=1\label{constraints two higgs
4}\,\,,
\end{align}
which generate only the lepton Yukawa couplings $L {H^I_d}{E_R}$ and
${L} {H^I_u} {N_R}$, but not any of the quark Yukawa couplings.

The analysis of the lepton and quark Yukawa couplings is very
similar and all results apply for this setup as well. As previously,
$E2_1$ and $E2_2$ give rise to a  $\mu$-term there
\begin{align}
\mu_{IJ} \, H^I_d \,H^J_u
\end{align}
for the Higgs fields. In opposite to triangle quiver discussed in
section \ref{setup 2}, the $\mu$-term is a $2\times 2$ matrix. In
case the instanton, inducing the missing couplings
\eqref{perturbative forbidden 2 Higgs}, gives rise to a factorizable
$\mu$-term matrix, only one Higgs pair receives mass of order
$(10^{13}-10^{16})GeV$, while the other one, which we identify with
the physical MSSM-Higgs pair, remains massless. Then potentially
another instanton, wrapping a larger cycle in the internal manifold
could generate  additional non-perturbative corrections to the
Yukawa-couplings and the $\mu$-term matrix. Due to the higher
suppression factor these corrections are negligible for all
couplings, apart for the mass term for the physical Higgs pair, for
which it gives the dominant contribution. In case the suppression
factor arising from the second instanton is of the order
$e^{-S^{cl}_{E2}}\sim 10^{-16}-10^{-15}$ the $\mu$-term is in the
desired range $(10^2-10^3) \,GeV$.

\begin{figure}[h]
\begin{center}
 \includegraphics[width=0.9\textwidth]{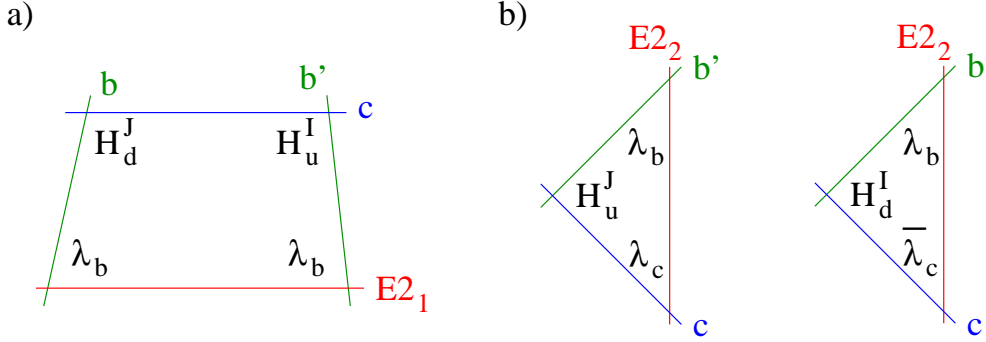}
\end{center}
\caption{\small Non-perturbative generation of $\mu$-term matrix
$\mu^{IJ}$.}\label{Higgssetuptwo}
\end{figure}

We now discuss under which circumstances the $2 \times 2$
matrix $\mu_{IJ}$ factorizes, thus leaving only one Higgs pair massless.
For an instanton $E2_1$ satisfying \eqref{constraints two higgs 1}
the charged zero modes $\overline{\lambda}_b$ get soaked by just one
disk diagram (see figure \ref{Higgssetuptwo}a)
\begin{align}
B^{IJ}=< \lambda_{b}\, H^I_d \, H^J_u \,  \lambda_{b}>
\end{align}
Here, there is no reason to expect a factorization of the matrix
$\mu_{IJ}\sim B^{IJ}$, thus both Higgs multiplets get a mass of the
order $e^{-S^{cl}_{E2_1}} \, M_s$. For the other class of
instantons, $E2_2$, satisfying \eqref{constraints two higgs 2}, the
four charged zero modes get soaked up by two disk diagrams (see
figure \ref{Higgssetuptwo}b)
\begin{align}
A^I_d=< \lambda_{b}\, H^I_d \, {\overline \lambda}_c> \qquad
\text{and} \qquad A^J_u=< \lambda_{b}\, H^J_u \, \lambda_c>.
\end{align}
Clearly, for this class of instantons the matrix $\mu_{IJ}\sim A^I_d
\, A^J_u$ factorizes, thus only one of the Higgs pairs acquires a
mass of the order $e^{-S^{cl}_{E2_2}} \, M_s$. The other, the
physical, Higgs pair remains massless. Thus, in a setup satisfying
\eqref{intersection number higgs2}, which exhibits both classes of
instantons, Yukawa couplings of the right order can be generated
without inducing a too large $\mu$-term for the physical Higgs pair.
The instanton $E2_2$ of the second type gives rise to the correct
texture of the Yukawa couplings, while the suppression factor,
$e^{-S^{cl}_{E2_1}}$, of the instanton $E2_1$ is of the right order
to induce a $\mu$-term for the physical Higgs pair.

The linear combination remaining light depends on the values of the
world-sheet amplitudes $A_d^I$, $A_u^J$. Note that now, unlike the
simplest triangle setup, the perturbative contribution to the
fermion mass matrices eq.(\ref{yukper}) has generically rank two,
since the massless physical field is a linear combination of two
fields with different worldsheet couplings.

For the local realization \eqref{configuration 2} only a
representant of the second type, satisfying \eqref{constraints two
higgs 2}, is present. It gives rise to all perturbatively missing
couplings listed in \eqref{perturbative forbidden 2 Higgs}. The
instanton $E2$ wraps the orientifold invariant cycle
\begin{align}
\pi_{E2} =&\,\frac{1}{4} \,[(1,0)(0,1)(0,-1)] + \frac{1}{4}\left(
\sum_{i,j \in (1,3)\times(1,2)} \Big[\alpha^{\theta}_{ij} \times
(0,-1)\Big]\right) \\ \nonumber +&\, \frac{1}{4} \left(\sum_{i,j \in
(1,2)\times(1,2)}\Big[ \alpha^{\theta'}_{ij} \times
(1,0)\Big]\right) + \frac{1}{4}\left( \sum_{i,j \in
(1,3)\times(1,2)} \Big[ \alpha^{\theta\theta'}_{ij} \times
[(0,1)\Big]\right) 
\,\,
\end{align}
and has the intersection pattern \eqref{constraints two higgs 2}.
For the analysis of the ${\cal N}=2$ sector we refer the reader to
the appendix \ref{appendix N=2}. To obtain realistic Yukawa
couplings the suppression factor should be in the range
$10^{-2}-10^{-5}$ which can be achieved, via \eqref{suppression1}
and \eqref{supprssion 2}, if the complex structure moduli $U_3=u$ is
stabilized in the range $(0.7-1)$. Here we assume again that the
$SU(3)$ gauge coupling is approximately $\alpha_a=1/24$ at the
string scale.

As discussed above the instanton $E2_2$ also induces a large
$2\times 2$ $\mu$-term matrix for the Higgs pairs, which factorizes.
Thus, only one Higgs pair gets supermassive, while the other one
remains massless. While in principle the $\mu$-term for the physical
Higgs pair can be generated by a different instanton, wrapping a
larger cycle in the internal manifold, this local example does not
exhibit such an instanton.

Analogous to the previous  setup the smallness of the Neutrino
masses could  be explained by the see-saw mechanism. The instanton
inducing Majorana mass term for the right-handed Neutrino has to
satisfy \eqref{majorana mass} and is expected to have a suppression
factor of the order $10^{-8}-10^{-5}$. For the local realization
\eqref{configuration 2} such an instanton is absent.

Summarizing we see that the triangle quiver with a doubled Higgs
sector can give rise to Yukawa couplings of the right texture and at
the same time also exhibit a $\mu$-term of the desired order. After
integrating out the superheavy Higgs pair the low effective action
resembles the one of the MSSM. Furthermore, in case the setup also
exhibits an instanton with the right suppression factor, satisfying
\eqref{majorana mass}, the see-saw mechanism can be realized, thus
the origin of the smallness of the Neutrino masses could  be
explained.

\section{Conclusion}

In this work we have analyzed MSSM-like quivers with respect to
their Yukawa couplings. Specifically we have focused on three
different configurations, each consisting of four different stacks
of D-branes, the \emph{square-quiver}, the \emph{triangle-quiver}
and the \emph{triangle-quiver with a doubled Higgs sector}. We have
shown that in principle in all three configurations all
perturbatively absent, but desired couplings could be generated via
stringy instantons.

Furthermore, we investigated the phenomenological implications of
these non-perturbative effects. The square-quiver, discussed in
section \ref{sqare quiver}, does not allow for a perturbative
realization of the Dirac Neutrino masses, thus the exponentially
suppressed non-perturbative generation for this Yukawa coupling
gives an intriguing explanation for the smallness of the Neutrino
masses. However, the square-quiver generically gives rise to a large
mixing between the first and third generation for the quark Yukawa
couplings, which requires a large fine-tuning to overcome.

For the triangle-quiver we show that such fine-tuning is not
necessary to match experimental observations. All perturbatively
absent couplings can potentially generated by a single instanton.
Clearly, in this case the neat mechanism for small Neutrino masses,
encountered for the square quiver does not apply. However, the
setup, realizing the triangle quiver, might in principle exhibit an
instanton inducing a large Majorana mass term for the right-handed
Neutrinos, thus realizing the see-saw mechanism. Nevertheless for
the local realizations of both triangle quivers no such instanton is
found. The simplest triangle quiver has a generic potential
drawback. The very same instanton generating all missing Yukawa
couplings, induces also a $\mu$-term for the Higgs pair, which is
far too large.

As we discuss in section \ref{setup 3}, if we allow for a doubled
Higgs sector then this phenomenological drawback might be
surmounted. While the analysis for the Yukawa-couplings is very
similar to the triangle quiver with just one Higgs pair, the
$\mu$-term arising from the Yukawa generating instanton is a $2
\times 2$ matrix. In case the instanton induced $\mu$-term matrix
factorizes only one of the Higgs pairs becomes supermassive while
the other one, which we identify with physical MSSM Higgs pair,
remains massless. A second instanton with a larger suppression
factor could then induce the $\mu$ term for the MSSM Higgs pair. In
section \ref{setup 3} we show that a setup, realizing the
triangle-quiver with a doubled Higgs sector, potentially exhibits a
class of instantons, generating all the missing Yukawa couplings and
giving rise a $\mu$-term matrix, which factorizes.

For both triangle quivers, discussed in section \ref{setup 2} and
\ref{setup 3}, we provide local realizations within the
$T^6/{\mathbb Z}_2 \times {\mathbb Z}_2'$ orientifold with torsion,
which gives rise to rigid cycles. We  show that they exhibit
instantons which generate the missing Yukawa couplings with the
right hierarchies. For the triangle quiver with a doubled Higgs
sector the Yukawa coupling inducing instanton gives rise to a
$\mu$-term matrix which factorizes. Thus for this local setup we can
realize all Yukawa coupling with the right hierarchies without
encountering any phenomenological drawbacks. However, this local
setup does not exhibit an instanton which could generate a
$\mu$-term of the order $(10^2-10^3)GeV$ for the MSSM Higgs pair.

We leave it for future work to extend the analysis to other quivers,
such as quivers based on three stacks of D-branes and quivers
leading to GUT-like spectrum. Moreover, it would be nice to find a
global realization of the discussed quivers, which satisfies the
severe constraints on the instanton zero mode structure. This seems
to be more promising in the T-dual Type I framework. Finally, it
would be interesting to perform a detailed analysis of the MSSM-like
Gepner orientifold constructions, provided in
\cite{Dijkstra:2004cc,Anastasopoulos:2006da}, with respect to their
Yukawa couplings.




\newpage

{\bf Acknowledgments}\\
We thank M. Ambroso, T. Brelidze,  M. Cveti{\v c}, I.
Garc\'ia-Etxebarria, C. Kokorelis, F. Marchesano, A. Uranga and T.
Weigand for useful discussions. This work has been supported by the
European Commission under RTN European Programs MRTN-CT-2004-503369,
MRTN-CT-2004-005105, by the CICYT (Spain) under project
FPA2006-01105, the Comunidad de Madrid under project HEPHACOS
P-ESP-00346 and the Ingenio 2010 CONSOLIDER program CPAN.

\newpage

\appendix

\section{$T^6/{\mathbb Z}_2 \times {\mathbb Z}_2'$ Orientifold \label{appendix orientifold} }

This appendix is dedicated to brief review of the $T^6/{\mathbb Z}_2
\times {\mathbb Z}_2'$ orientifold with Hodge numbers
$(h_{11},h_{12})=(3,51)$. We adopt the notation of
\cite{Blumenhagen:2005tn}, to which we refer the reader for further
details. The orbifold group consists of two generators $\theta$ and
$\theta'$ acting as reflection in the first two and last two tori,
respectively, while there combination $\theta\theta'$ amounts into a
reflection in the first and third torus.

As usual there are the bulk cycles
 \bea \Pi_a^B = 4 \, \bigotimes_{I=1}^3
\,(n_a^I [a^I] + \widetilde m_a^I [b^I]), \eea defined in terms of
the fundamental one-cycles $[a^I], [b^I]$ of the $I$-th $T^2$ and
the corresponding wrapping numbers $n_a^I$ and $\widetilde m_a^I=
m_a^I + \beta^I n_a^I$. Here $\beta^I$ takes the value $0$ and $1/2$
for rectangular and tilted tori, respectively.

In addition this backgrounds gives rise to another class of cycles,
the so called $g$-twisted cycles. All three actions, $\theta$,
$\theta'$ and $\theta\theta'$ exhibit 16 fixed points which after
blowing up give rise to two-cycles with the topology of ${\mathbb
P}_1$. In combination with the fundamental one-cycle invariant under
the respective action they construct the $g$-twisted cycles \bea
\Pi^g_{ij} = \Big[\alpha^g_{ij} \times
(n^{I_g},\widetilde{m}^{I_g})\Big]. \eea Here ${i,j} \in \{1,2,3,4\}
\times \{1,2,3,4\}$ labels one of the 16 blown-up fixed points of
the orbifold element $g  = \theta, \theta', \theta \theta' \in
\Z_2\times \Z_2'$ and $I_g$ denotes the $g$-invariant one-cycle with
$I_g = 3,1,2$ for $g  = \theta, \theta', \theta \theta'$.

Rigid cycles are charged under all three sectors $\theta$, $\theta'$
and $\theta\theta'$ and take the form \bea \Pi^F = \frac{1}{4} \Pi^B
+ \frac{1}{4} \Bigl( \sum_{i,j \in S_{\theta}}
\epsilon^{\theta}_{ij} \Pi^{\theta}_{ij} \Bigr)+ \frac{1}{4} \Bigl(
\sum_{j,k \in S_{\theta'}} \epsilon^{\theta'}_{jk}
\Pi^{\theta'}_{jk}  \Bigr) + \frac{1}{4} \Bigl(  \sum_{i,k \in
S_{\theta \theta'}} \epsilon^{\theta \theta'}_{ik} \Pi^{\theta
\theta'}_{ik}  \Bigr), \eea where $S_g$ is the set of fixed points
in the $g$-twisted sector. The $\epsilon^g_{ij}=\pm 1$ correspond to
the two different orientation the brane can wrap the blown up
${\mathbb P}_1$ and are subject to various consistency conditions
\cite{Blumenhagen:2005tn}.

The orientifold action $\Omega\mathcal{R}$ for the bulk cycles takes
the usual form
\begin{align}\label{orientifold action1}
\Omega\mathcal{R}:
[(n_1,\widetilde{m}_1)(n_2,\widetilde{m}_2)(n_3,\widetilde{m}_3)]\rightarrow
[(n_1,-\widetilde{m}_1)(n_2,-\widetilde{m}_2)(n_3,-\widetilde{m}_3)].
\end{align}
For the $g$-twisted cycle $\Omega\mathcal{R}$ acts
\begin{align}
\label{Omegatwisted} \Omega\mathcal{R}: \,\Big[\alpha^g_{ij}\times
(n^{I_g},\widetilde{m}^{I_g})\Big]\rightarrow
\eta_{\Omega\mathcal{R}}\,\eta_{\Omega\mathcal{R}g} \Big[
\alpha^g_{\mathcal{R}(i)\mathcal{R}(j)}
\times(-n^{I_g},\widetilde{m}^{I_g})\Big],
\end{align}
where $\eta_{\Omega\mathcal{R}g}=\pm 1$ denote the orientifold
charges of the different sectors and are subject to the constraint
\begin{align}
\label{etaconst}
\eta_{\Omega\mathcal{R}}\,\eta_{\Omega\mathcal{R}\theta}\,
\eta_{\Omega\mathcal{R}\theta'}\,\eta_{\Omega\mathcal{R}\theta\theta'}=-1.
\end{align}
The reflection $\mathcal{R}$ leaves all fixed points of an untilted
two-torus invariant and acts on the fixed points in a tilted
two-torus as
\begin{align}
\mathcal{R}(1)=1, \qquad \mathcal{R}(2)=2, \qquad\mathcal{R}(3)=4,
\qquad\mathcal{R}(4)=3.
\end{align}

With the orientifold action \eqref{orientifold action1} the fixed
point locus is
\begin{align*}\pi_{O6}=\,&
2^3\, \eta_{\Omega\mathcal{R}}\, [(1,0)(1,0)(1,0)]
+2^{3-2\beta_1-2\beta_2} \,\eta_{\Omega\mathcal{R}\theta}\,
[(0,1)(0,-1)(1,0)]\\ \\& +2^{3-2\beta_2-2\beta_3}
\,\eta_{\Omega\mathcal{R}\theta'}\, [(1,0)(0,1)(0,-1)]+ 2^{ 3-
2\beta_1 -2\beta_3} \,\eta_{\Omega\mathcal{R}\theta\theta'}\,
[(0,-1)(1,0)(0,1)]
\end{align*}

\begin{table}
\centering
\begin{tabular}{|c|c|}
\hline
Representation  & Multiplicity \\
\hline $ \Yasymm_a$
 & ${1\over 2}\left(\pi_a\circ \pi'_a+\pi_a \circ  \pi_{{\rm O}6} \right)$  \\
$\Ysymm_a$
     & ${1\over 2}\left(\pi_a\circ \pi'_a-\pi_a \circ  \pi_{{\rm O}6} \right)$   \\
$( a,{\overline b})$
 & $\pi_a\circ \pi_{b}$   \\
 $(a,b)$
 & $\pi_a\circ \pi'_{b}$
\\
\hline
\end{tabular}
\vspace{2mm} \caption{Chiral spectrum for intersecting D6-branes.}
\label{table chiral spectrum}
\end{table}

Generically, the chiral spectrum is given by the topological
intersection numbers displayed in table \ref{table chiral spectrum}
\footnote{Note that we choose the convention that positive
intersection number $\pi_a \circ \pi_b$ corresponds to matter
transforming as $(a,{\overline b})$.}. Given two branes $a$ and $b$
in the $T^6/{\mathbb Z}_2 \times {\mathbb Z}_2'$ orbifold
background, the topological intersection numbers for the bulk part
is
\begin{align}
 \pi^B_a \circ\pi^B_b = 4 \prod^3_{i=1} (n^i_a
\widetilde{m}^i_b-n^i_b \widetilde{m}^i_a)\label{intersection
formula bulk}
\end{align}
and for the twisted sector takes the form
\begin{align}
 \Big[\alpha^{g}_{ij}\times (n^{I_g}_a,\widetilde{m}^{I_g}_a) \Big]\circ \Big[\alpha^{h}_{kl} \times (n^{I_h}_b,\widetilde{m}^{I_h}_b)\Big] =
 4 \,\delta_{ik} \,\delta_{jl}\, \delta^{gh} \,(n^{I_g}_a \, \widetilde{m}^{I_h}_b - n^{I_h}_b \,
 \widetilde{m}^{I_g}_a)\,\,.
 \label{intersection formula twisted}
\end{align}
In both local realizations presented in section \ref{setup 2} and
\ref{setup 3} the matter $D6$-branes wrap fractional cycles, cycles
which are charged only under one twisted sector $g$
\begin{align}
 \pi^F= \frac{1}{2} \pi^B +
\frac{1}{2} \left( \sum_{i,j \in S_{g}} \epsilon^{g}_{ij}
  \Big[\alpha^{g}_{ij}\times (n^{I_g}_a,\widetilde{m}^{I_g}_a)
  \Big]\right)\,\,.
  \label{intersection number fractional}
\end{align}
This class of cycles are only rigid in two tori and can move freely
in the torus invariant under the action $g$. Then, for two
fractional branes charged under a different twisted sector
 the intersection number is simply given by
 \begin{align}
\pi^F_a \circ \pi^F_b =\prod^3_{i=1} (n^i_a \widetilde{m}^i_b-n^i_b
\widetilde{m}^i_a)\label{intersection fractional brane}
\end{align}

To achieve global consistency the RR charges of the branes and
orientifolds have to be cancelled\footnote{Moreover, there are
K-theory constraints which have to be satisfied.}. That translates
into conditions on the bulk and twisted sector of the branes. Since
we are only interested in local constructions we ignore them and
refer the interested reader to \cite{Blumenhagen:2005tn}.

Finally, to ensure supersymmetry all branes have to be aligned to
the orientifold plane. This amounts into two constraints each brane
has to satisfy. Expressed in terms of the wrapping numbers they are
\begin{align}
{\widetilde m}^1\,{\widetilde m}^2\,{\widetilde m}^3-\sum_{I\neq
J\neq K}\frac{n^I\,n^J\, {\widetilde m}^K}{U^{I}\,U^{J}}=0
\label{susy1}
\end{align}
and
\begin{align}
n^1\,n^2\,n^3-\sum_{I\neq J\neq K}{\widetilde m}^I\,{\widetilde
m}^J\, n^K\,U^{I}\,U^{J}>0\,\,, \label{susy2}
\end{align}
where $U^I$ denotes the complex structure modulus $U^I=R^I_Y/R^I_X$
of the $I-th$ torus with radii $R^I_X$ and $R^I_Y$.


\section{${\cal N}=2$ instanton zero mode sector  \label{appendix N=2}}
This appendix deals with the detailed analysis of the ${\cal N}=2$
instantonic zero modes arising in the examples presented in section
\ref{setup 2} and \ref{setup 3}. We follow closely the analysis
performed in chapter 5 of \cite{Blumenhagen:2005tn}.

\begin{figure}[h]
\begin{center}
 \includegraphics[width=1.0\textwidth]{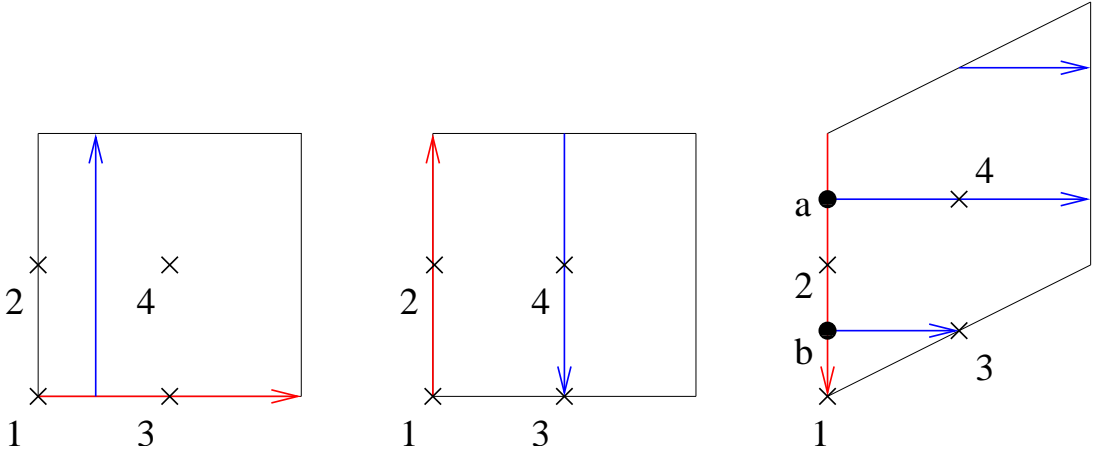}
\end{center}
\caption{\small Brane $c$ (blue) and the instanton $E2_1$
(red).}\label{intersection E2 c for instanton 1}
\end{figure}

Consider the four stack model giving rise to just one Higgs pair
discussed in section \ref{setup 2}. The $D6$-branes configuration
exhibits two different instantons $E2_1$ and $E2_2$ satisfying the
two intersection pattern \bea
&& I_{E2_1b}=-1 \qquad I_{E2_1a}=I_{E2_1c}=I_{E2_1d}=0 \label{instanton1app} \\
&&I_{E2_2b}=-1 \qquad I_{E2_2a}=I_{E2_2c}=I_{E2_2d}=0 \qquad
I^{{\cal N}=2}_{E2_1c}=1 \label{instanton2app}\,\,. \eea For $E2_1$
there are no massless vector-like states between the $D6$-branes and
the instanton, while for $E2_2$ there is exactly one massless
vectorlike state arising at the intersection of the instanton $E2_2$
and the $D6$-brane $c$.

Let us start by analyzing the ${\cal N}=2$ sector for the instanton
$E2_1$, whereas we focus on the $E2-c$ sector. An analogous
discussion applies for all the other $E2_1-D6$ sectors. Figure
\ref{intersection E2 c for instanton 1} depicts the intersection
between the instanton $E2_1$ and D-brane $c$. Note, that the two
cycles  $D6_c$ and $E2_1$ wrap (see equations \eqref{configuration
1} and \eqref{instanton 1 setup 1}), are parallel in the second
torus, but go through different fixed points. Thus both cycles are
separated in the second torus, the ${\cal N}=2$ modes are massive
and one does not observe any massless vectorlike states in
$E2_1-D6_c$ sector. A similar analysis reveals that there are no
vectorlike states in any of the other $E2_1-D6$ sectors.

For the instanton $E2_2$ similar arguments explain the absence of
massless vectorlike states for all $E2_2-D6$ sectors apart from the
$E2_2-c$ sector. In opposite to $E2_1$, $E2_2$ and $c$ are not
separated in the second torus, but both $E2_2$ and $c$ pass through
the same fixed points (see figure \ref{intersection E2 c for
instanton 2 }). Thus the ${\cal N}=2$ sector does not get massive
and one potentially observes massless vectorlike states.

\begin{figure}[h]
\begin{center}
 \includegraphics[width=1.0\textwidth]{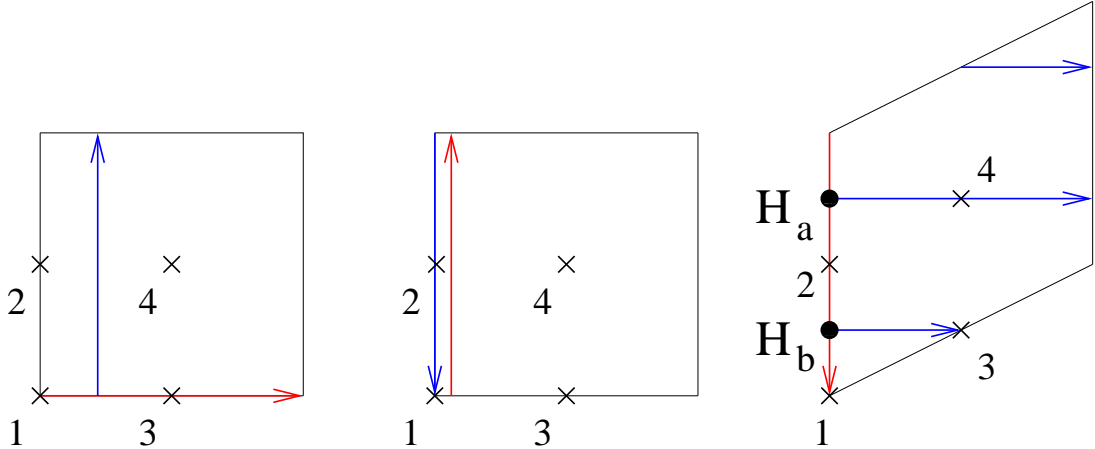}
\end{center}
\caption{\small Brane $c$ (blue) and the instanton $E2_2$
(red).}\label{intersection E2 c for instanton 2 }
\end{figure}
Subsequently we analyze how many of them are present. The bulk
sector implies the presence of 2 hypermultiplets in the ambient
space\footnote{The number of hypermultiplets is given by the number
of intersections in the first and third torus $I^{{\cal
N}=2}_{{E2_2c}^{\text{bulk}}}= (n^1_{E2_2}\,m^1_c-n^1_c\,
m^1_{E2_2})(n^3_{E2_2}\,m^3_c-n^3_c\, m^3_{E2_2})$ .}, which we
denote by $H_a$ and $H_b$ (see figure \ref{intersection E2 c for
instanton 2 }). Including the orbifold action, which acts as
\begin{align}\nonumber
\theta:    \qquad &(H_a, H_b) \rightarrow (H_a, H_b)\\
\theta':     \qquad&(H_a, H_b) \rightarrow (H_b, H_a)\\ \nonumber
\theta\theta': \qquad &   (H_a, H_b) \rightarrow (H_b, H_a)
\end{align}
we see that only the combination $H_a + H_b$ survives. Thus there is
indeed exactly one charged vectorlike pair of zero modes,
$\lambda_c$ and ${\overline \lambda}_c$, in the $E2_2-c$ sector.

\begin{figure}[h]
\begin{center}
 \includegraphics[width=1.0\textwidth]{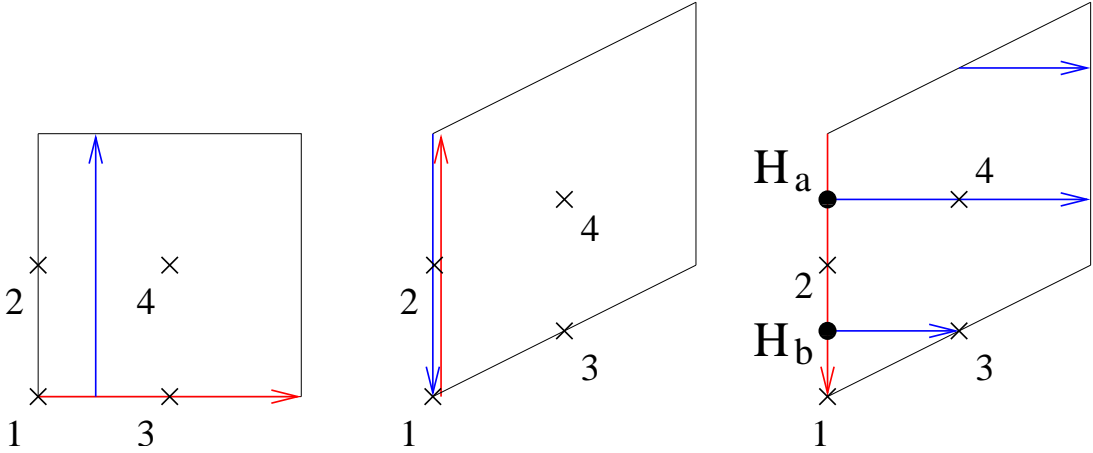}
\end{center}
\caption{\small Brane $c$ (blue) and the instanton $E2$
(red).}\label{fig intersection E2 c for two higgs  }
\end{figure}

Let us turn to the example presented in section \ref{setup 3}. This
D6-brane configuration comprises an instanton with the intersection
pattern
\begin{align}
I_{E2_2b}=-1\qquad I_{E2_2a}= I_{E2_2c}=I_{E2_2d}=0\qquad I^{{\cal
N}=2}_{E2_2c}=1 \label{intersection E2 c for two higgs setup}\,\,.
\end{align}

Analogously to the setup above there are no massless vectorlike
states between the instanton $E2$ and $D6$-brane $a$, $b$ and $d$.
Figure \ref{fig intersection E2 c for two higgs  } displays the
intersection between the cycles the instanton $E2$ and the D-brane
$c$ wrap. They lie on top of each other in the second torus, giving
potentially massless ${\cal N}=2$ states in the $E2-c$ sector. As
previously, in the ambient space we have 2 hypermultiplets, which
are subject to the orbifold action. A similar analysis as above
reveals that only one combination, $H_a+H_b$ survives this action.
Thus as desired we have exactly 2 instanton zero modes charged under
the global $U(1)_c$, namely $\lambda_c$ and ${\overline \lambda}_c$.

\clearpage \nocite{*}
\bibliography{rev}
\bibliographystyle{utphys}

\end{document}